\documentclass[fleqn,usenatbib]{mnras}
\usepackage{newtxtext,newtxmath}

\usepackage[T1]{fontenc}
\usepackage{ae,aecompl}
\usepackage{dcolumn}
\usepackage{bm}

\usepackage{psfig}
\usepackage{graphicx}  
\usepackage{mnras_cite}  
\usepackage{xcolor}
\usepackage{hyperref}
\newcommand{\bea}{\begin{eqnarray}}
\newcommand{\eea}{\end{eqnarray}}
\newcommand{\blue}{\color{black}}

\newcommand{\beq}{\begin{equation}}
\newcommand{\eeq}{\end{equation}}

\newcommand{\beqa}{\begin{eqnarray}}
\newcommand{\eeqa}{\end{eqnarray}}

\def\la{\mathrel{\mathpalette\fun <}}
\def\ga{\mathrel{\mathpalette\fun >}}
\def\fun#1#2{\lower3.6pt\vbox{\baselineskip0pt\lineskip.9pt
\ialign{$\mathsurround=0pt#1\hfil##\hfil$\crcr#2\crcr\sim\crcr}}}

\begin{document}

\title[Blending and obscuration in weak lensing magnification]
{Blending and obscuration in weak lensing magnification}

\author[E.Gaztanaga, S.J.Schmidt, M.D.Schneider, J.A.Tyson]
{\parbox{\textwidth}
{
E.~Gaztanaga$^{1,2}$, S.~J.~Schmidt$^3$, M.~D.~Schneider$^4$  \& J.~A.~Tyson$^3$ \\
\small{
$^1$ Institute of Space Sciences (ICE, CSIC), Campus UAB, Carrer de Can Magrans, s/n,  08193 Barcelona, Spain\\
$^2$ Institut d'Estudis Espacials de Catalunya (IEEC), E-08034 Barcelona, Spain  \\
$^3$ Department of Physics, University of California, Davis, One Shields Avenue, Davis, CA 95616, USA \\
$^4$  Lawrence Livermore National Laboratory, P.O. Box 808 L-211, Livermore, CA 94551, USA\\
}}}

\maketitle

\begin{abstract}
We test the impact of some systematic errors in weak lensing 
magnification measurements with the COSMOS 30-band photo-$z$ Survey flux limited to $I_{auto}<25.0$ using correlations of both source galaxy counts and magnitudes. 
Systematic obscuration effects are measured by comparing counts and magnification correlations. 
We use the ACS-HST catalogs to identify potential blending objects (close pairs) and perform the magnification analyses with and without blended objects.
We find that blending effects start to be important ($\sim$ 0.04~mag obscuration) at angular scales smaller than 0.1 arcmin. Extinction and  other systematic obscuration effects can be as large as 0.10~mag (U-band) but are typically smaller than 0.02~mag depending on the band. After applying these corrections,
we measure a $3.9\sigma$ magnification signal that is consistent for both counts and magnitudes. The corresponding projected mass profiles of galaxies at redshift  $z \simeq 0.6$ ($M_I \simeq -21$) is $\Sigma= 25\pm 6 M_{\sun}h^3/pc^2$ 
at 0.1 Mpc/h, consistent with  NFW type profile with  $M_{200} \simeq 2 \times 10^{12} M_{\sun} h/pc^2$.
Tangential shear and flux-size magnification over the same lenses show similar mass profiles.
We conclude that magnification from counts and fluxes 
using photometric redshifts has the potential to provide complementary weak lensing information in future wide field surveys once we carefully take into account systematic effects, such as obscuration and blending.

\end{abstract}

\maketitle

\section{Introduction}\label{sec:intro}

Gravitational lensing  can be used to measure the total matter distribution projected onto the sky (for a review see Bartelmann \& Schneider 2001) and in combination with galaxy clustering provides a powerful tool to study cosmology and structure formation (see e.g. Bernstein \& Cai 2011, Cai \& Bernstein 2012, Gaztanaga et al.~2012, Weinberg et al. 2013, Eriksen \& Gaztanaga 2015a-c,  Eriksen \& Gaztanaga 2018).

In the weak lensing regime gravitational distortions result in shear and magnification in the shapes, sizes, fluxes, and counts 
of galaxies in the background of the lenses we want to study. 
{\blue  Magnification changes the observed galaxy fluxes and this manifests itself in two separate observational quantities: galaxy number counts   (e.g. see  Schneider 1987, Narayan 1989)
and magnitude-counts cross-correlations (e.g. see Menard et al. 2010). In this paper we will focus on these two effects and will compare them to magnification from galaxy sizes and shear derived from shapes using the same galaxy sample (Schmidt et al. 2012). 
Van Waerbeke (2010) discusses the cosmological gain from combining magnification and shear.}

The magnification signal from cross-correlation of galaxy samples is small  (a per cent level change in clustering on arcminute scales) while the corresponding statistical noise  (from the auto-correlation of the samples) can be large. We therefore need either a large area or large density to sample enough galaxy pairs to achieve a significant signal-to-noise ratio (SNR) magnification measurement. These two situations are complementary because the survey cost is proportional to the total number of objects and therefore the number of pairs available. The two common strategies are to use either low densities with large survey area or high densities with small survey area.  The former samples large scales (mainly probing the 2-halo term) while the latter is usually restricted to small scales  (probing the 1-halo term).
{\blue Besides SNR we also need accuracy of the measurements and there are several systematic bias that can affect such small cross-correlation signal (see Hildebrandt et al. 2016).}

Current magnification measurements have been achieved over dilute spectroscopic samples spread over large areas (e.~g.~Scranton et al.~2005, Bauer et al.~2014). Samples are  well separated in redshifts by selecting foreground spectroscopic subsamples (e.~g.~LRG or clusters) which are cross-correlated with high redshift selected samples: spectroscopic QSO (e.~g.~ Gaztanaga 2003, Menard et al. 2010) or photometric Dropout Galaxies (e.~g.~Hildebrant et al. 2009, Morrison et al. 2012).
These measurements therefore sample large (2-halo terms) correlations.
The small scale weak lensing regime (of 1-halo term) is currently better explored with sheared shape measurements over smaller and denser samples using the  Hubble Space Telescope (HST) Advanced Camera for Surveys (ACS) (e.~g.~Leuthaud et al. 2007 and references therein) or larger ground based 
WL Surveys such as Deep Lens Survey (Wittman et al.~2002, Jee et al. 2013),  CFHTLenS (Heymans et al.~2013, {\blue Ford et al 2015}),  KIDS (Hildebrandt et al.~2017), DES (Abbott et al.~2018), or HSC (Hikage et al.~2019).

Schmidt et al.~(2012) presented one of the few existing comparisons of magnification and shear measurements using the same lens samples. This work uses the same COSMOS reference catalog as the one we use in our analysis. The main difference with Schmidt et al. (2012) is that we use flux limited samples (rather than galaxy groups) for lenses, and that we estimate magnification from the combination of flux and (HST) sizes, while here we focus on magnification from counts and fluxes. The motivation for our approach is to test systematic effects in magnification measurements of deep samples where morphological (space based) information is not available or is not that reliable. We will also compare our results with those using shear and flux-size magnification for the same lens samples.

Future deep photometric surveys, such as the Legacy Survey of Space and Time (LSST) to be carried out in 2022-2032 by the the Rubin Observatory,
Euclid, and WFIRST, have the potential to measure magnification for flux limited samples (e.g.~the LSST Science Book). For deep samples ($I_{auto}>25$) ground based shape measurements are limited by systematics, and it might be harder to measure weak lensing from shapes and sizes than from flux and counts, as presented in this paper. But there are several systematic effects that could limit the potential of such magnification measurements. Among them are: 1) the accuracy of the photometric redshift selection and characterization (which could add  intrinsic cross-correlation)  2) flux/count obscuration effects due to the atmosphere, extinction or data reduction (e.g background subtraction) 3) blending of sources, which can affect both the counts and the fluxes. In this paper we explore these three effects in a working example, the COSMOS field which has a wealth of deep and accurate data, including space based observations, to explore some of these issues. 

This paper is organized as follows. In \S 2 we present the modelling of the magnification, obscuration and blending effects.
In \S 3 we present the data selection and validation. 
In \S 4 we present the magnificaiton results to finish with the conclusions in \S 5.

\section{Modeling}

In this section we discuss how magnification can be measured using either galaxy counts or fluxes.
We also present a method to combine cross-correlation measurements of density fluctuations (counts) and magnitudes (or fluxes) to account for systematic effects. 
\S\ref{sec:mag} introduces the definition of magnification and shows its effects on individually observed magnitudes. \S\ref{sec:obs} adds the impact of obscuration to observed magnitudes, while \S\ref{sec:cou} shows how counts can be affected by magnification and obscuration. The effects of blending are presented in \S\ref{sec:ble}.  \S\ref{sec:sep} proposes a way to separate the effects of magnification and obscuration, while  \S\ref{ssec:cross} introduces the cross-correlation measurements that we use in this paper.

\subsection{Magnification}
\label{sec:mag}

Magnification of flux  $\mu$ in gravitational lensing distortion is:
\beq
\mu \equiv 1+\delta_\mu = {1\over{det A}} = {1\over{(1-\kappa)^2-|\gamma|^2}}
\label{eq:mu}
\eeq
where $A$ is the Jacobian matrix for the lensing transformation
(see e.g. Bartelmann \& Schneider 2000). 
In the weak lensing (WL) regime, fluctuations in magnification $\delta_\mu$, 
convergence $\kappa$ and shear $\gamma$ are closely related. 
Small fluctuations in convergence  $\delta_\kappa$,
according to Eq.\ref{eq:mu},  are half as large as magnification
$\delta_\mu=2\delta_\kappa$.

The measured flux $F$ is changed relative to the emitted
flux $F_0$ as $F=F_0 \mu$.  So the change to the apparent magnitude
{\blue in band $\lambda$ is  :
\beq
\delta m_\lambda =  - {2.5\over{\ln{10}}}\delta_\mu
\label{eq:deltam0}
\eeq
which is independent of the observational wavelength $\lambda$ used to measure the flux}.  This independence is helpful in separating the magnification signal from systematic effects that are chromatic, such as obscuration. 

\subsection{Magnitudes and Obscuration}
\label{sec:obs}

In addition to magnification, observed magnitudes can be biased by galactic and extra galactic extinction/absorption, optical and atmospheric variations, blending and  errors in the process of 
data imaging reduction (e.g. see for example Chang et al. 2015). From now on we will refer to the ensemble of all these effects as ``obscuration" and we will not attempt to disentangle the different possible origins of these contributions. However, we will attempt to 
disentangle them from magnification by comparing their impact on magnitudes and counts.

{\blue
The measured flux $F$ is changed as $F=F_0 \mu e^{-\tau_{\lambda}}$ where $\tau_{\lambda}$ is
the optical depth for obscuration effects
and $F_0$ is the flux in the absence of obscuration.
So the total combined change in observed magnitude due to magnification and obscuration is then:
\beq
\delta  m_\lambda = {2.5\over{\ln{10}}} \left( \tau_{\lambda} -\delta_\mu \right) =  - {2.5\over{\ln{10}}}\delta_\mu + A_\lambda
\label{eq:deltam}
\eeq
where $A_\lambda$ is the magnitude of obscuration, which could depend on wavelength $\lambda$ (contrary to magnification).

The change in a magnitude of wavelength $\lambda$ for an individual galaxy due to lensing  alone: $\delta m= - {2.5\over{\ln{10}}} \delta_\mu$ is the same for all $\lambda$. For a sample of galaxies,
the corresponding change for the magnitude difference: 
\bea
\Delta_\lambda
&\equiv& \delta ( m_\lambda -\bar{m}_\lambda)  =  \delta  m_\lambda -\delta \bar{m}_\lambda 
\label{eq:delatL} \\ \nonumber
\bar{m}_\lambda &\equiv& <m_\lambda>
\eea
 depends on how the mean
 $<m_\lambda>$ changes due to magnification.
 This is a function of the distribution of magnitudes and the way that the  galaxies are selected. 
If galaxies are selected based on magnitude, the effects of magnification could move galaxies inside or outside the sample selection limits, which we denote here by $m_*$.
The change in the survey magnitude limit $m^*_\lambda$ due to magnification
is: $\delta m^*_\lambda=-\delta m_\lambda={2.5\over{\ln{10}}} \delta_{\mu}$, so:
\beq
\delta \bar{m}_\lambda ={\partial m^*_\lambda\over{\partial\mu}} {\partial\bar{m}_\lambda\over{\partial m^*_\lambda}} ~ \delta_\mu
= {2.5\over{\ln{10}}} ~ {\partial \bar{m}_\lambda\over{\partial m^*_\lambda}} ~ \delta_\mu
\eeq
so that:
\beq
\Delta_\lambda = -{2.5\over{\ln{10}}} ~\left( 1+ {\partial \bar{m_\lambda}[m^*_\lambda]\over{\partial m^*_\lambda}}
\right)  \delta_\mu
\equiv \alpha_\lambda~\delta_\mu
\label{Eq:mag}
\eeq
where $\alpha_\lambda$ accounts for both the changes in individual magnitudes and
the changes induced by the selection. This is the equivalent of
$C_s$ in Eq.12 of Menard et al. (2010). We have unified the notation here so that the global magnification factors for counts (see below) and magnitudes are called $\alpha$ with different subscripts. However, we note that this notation varies in the literature.}
If we now add the effect of
obscuration, we have:
\beq
\Delta_\lambda = \alpha_\lambda~\delta_\mu+ A_\lambda - \left({\ln{10}\over{2.5}} \alpha_\lambda+1\right) A_\lambda*
\label{Eq:mag2}
\eeq
as obscuration also moves galaxies across the sample selection limits.
Here $A_\lambda*$ refers to the obscuration in the magnitude band
$\lambda*$ used to select galaxies, while $A_{\lambda}$ is the
obscuration for the magnitude band $\lambda$ that we use as data
vector to measure magnification. In the case considered in this paper, 
both magnitude bands are the same, and the above equations 
simplifies a bit:
\beq
\Delta_\lambda = \alpha_\lambda~\delta_\mu- {\ln{10}\over{2.5}} \alpha_\lambda A_\lambda
\label{Eq:mag3}
\eeq

\subsection{Galaxy counts}
\label{sec:cou}

Lensing does not change surface brightness of the source, but
it changes the area of the background
sources behind lenses, which induces a fluctuation
in the galaxy counts: 

\beq
\delta_G \equiv {dN\over{N}} = -\delta_\mu.
\label{eq:deltaG}
\eeq

The change in individual source magnitudes can 
 also induce an additional change to $\delta_G $ if there is a shift in the mean
 survey magnitude limit $m_*$, where the magnification boosts faint, previously undetected galaxies, above the background detection threshold, i.~e.~
$dm^*_\lambda=-dm_\lambda={2.5\over{\ln{10}}} \delta_{\mu}$, so that:
\beq
\delta_G  = {1 \over{N}}  {dN \over{dm^*_\lambda}}  dm^*_\lambda =
 2.5 {d\log_{10} N\over{dm^*_\lambda}} \delta_{\mu} 
\eeq

Adding both contributions gives
\beq
\delta_G =  \left( 2.5 {d\log_{10}{N}\over{dm^*_\lambda}} -1 \right) \delta_\mu \equiv \alpha_c \delta_\mu
\label{Eq:counts}
\eeq
{\blue This notation is equivalent of $\alpha_c=\alpha-1$ in Eq.4 of Scranton et al. 2005.}
Thus the sign (and amplitude) of the magnification effect depends on the slope of the observed number counts $\alpha_c$. This will be tested later on in section 3.2.

As shown in Eq.~\ref{Eq:mag2},
obscuration will contribute to the change in the counts through the change of $A_\lambda^*$ in the magnitude limit $m^*_\lambda$. So the total contribution to $\delta_G$ is:

\beq
\delta_G = \alpha_c \delta_\mu - {\ln{10}\over{2.5}} (\alpha_c+1) A_\lambda^*
\label{Eq:counts2a}
\eeq
Note that for $\alpha_c \simeq -1$, we have $\delta_G \simeq -\delta_\mu$ and the signal is independent of obscuration.

\begin{figure}
\centering  \includegraphics[width=3.3in]{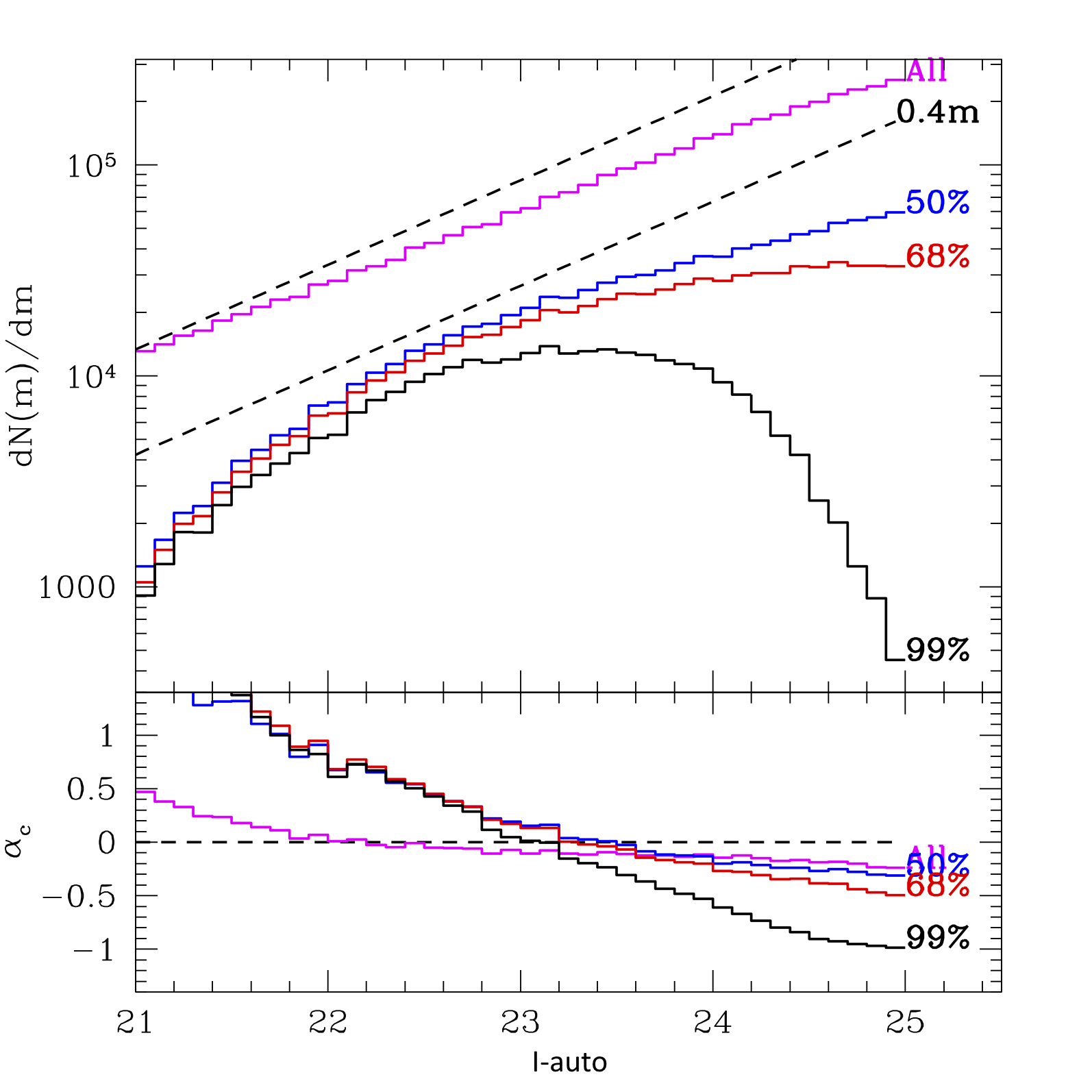}
\caption[]{Differential number counts $dN/dm$ as a function of  $I_{auto}-auto$ band (top panel) and logarithmic count slope  $\alpha_c$ (bottom) for all galaxies (pink)
and for galaxies selected with 50\% (blue), 68\% (red) or
99\% (black) photo-$z$ percentiles inside photo-$z$ bin $0.8<z<1.2$. In this paper we will use the 99\% selection at $I_{auto}<25$ which gives a total surface density of $\simeq 4$ galaxies/arcmin$^2$. While the photo-$z$ cut reduces the sample size (by a factor of 13), the reduction is necessary to avoid contamination between the lens and source bins.  The cut also produces a slope of $\alpha_C \sim -1$ at $I_{auto}<25$ sample, which maximizes the signal and minimizes the effect of the second term in Eq.12.}
\label{fig:dNdm-i}
\end{figure}

Fig.~\ref{fig:dNdm-i} shows the differential number counts  $dN/dm_\lambda$ as a function of the $m_\lambda=I_{auto}$ magnitude 
for the full source sample (labeled "All" in pink) and for different photo-$z$ selections. In each selection we only 
include galaxies which have $50\%$, $68\%$ or $90\%$  probability of being within $0.8<z<1.2$ (according to the  probability distribution provided in the photo-$z$ catalog for each galaxy).
The bottom panel shows $\alpha_c$ defined in Eq.~\ref{Eq:counts} above. In this paper we use the 99\% cut 
to reduce the contamination source and lens bins.


Fainter objects have larger photometric uncertainties and less certain photo-$z$ estimates, thus the photo-$z$ quality cuts reduce the number
of faint objects in the sample, thus the more stringent the quality cut, the more galaxies are removed, and preferentially removed at fainter magnitudes, as can be seen in Fig.~1.  These cuts can produce a number counts slope consistent with zero, which results in $\alpha_c \rightarrow -1$ in Eq.~\ref{Eq:counts}. This has several advantages: it reduces the uncertainty in the $\alpha_c$ estimation, it increases the signal of the magnification effect (note how the 50\% and 68\% cuts with slopes closer to zero) and minimizes the impact of the second term in Eq. 12. That the magnification signal is expected to be negative also helps to separate cross-correlation from magnification from the one from photo-$z$ contamination, which always gives positive signal.

\subsection{The effects of Blending in the counts}
\label{sec:ble}

{\blue At total apparent magnitude  in the I-band $I_{auto}<25$}
the size and density of galaxies is such that there is a small, but non negligible, chance for two galaxies to appear as a blended object in {\blue space} based observations (see Dawson et al.~2016).
Data reduction software and background estimation can also blend or split images.  These blending effects can also change the measured counts and the fluxes. The change in the fluxes is already accounted for under the obscuration effect (\S 2.2),  which includes any changes in fluxes due to the treatment of blending and background subtraction in the image reduction. {\blue
On top this, blending can also reduce the galaxy counts.}
Dawson et al.~(2016) compare HST and Subaru
imaging  over the COSMOS field ({\blue using a parent sample nearly identical to the one used in our analysis}) to estimate the effects of blending in the observed local projected density of galaxies  $n_G$ to find:
\beq
\Delta n_B = - B n_G^{3.15} 
\label{eq:dn}
\eeq
with $B \simeq 2.3 \times 10^{-6} arcmin^{-1}$ where $n_G$ is the true
(intrinsic) total density at a given sky position (without
magnification or obscuration). 
We have tested that the above expression works well for our
samples. For that test we used the COSMOS  ACS sample (Leauthaud et al.~2007)  to calculate the
distribution of number of pairs of galaxies as a function of the
background density.  The fraction of objects separated by less than 0.5
to 1.0 arcsec, which is the typical size of ACS galaxies+Subaru
seeing and are therefore potential blends, 
agrees well with the above relation for the densities $n_G$ in our samples.

Blending can therefore  reduce the background fluctuation by $\delta_G^B \simeq -B n_G^{2.15}$.
Including all effects:
\beq
\delta_G = \alpha_c \delta_\mu - {\ln{10}\over{2.5}} (\alpha_c+1) A_\lambda^*  -B n_G^{2.15} 
\label{Eq:counts2}
\eeq

In our analysis, samples are restricted to $I_{auto}<25$ where $\bar{n}_G \simeq 40$
galaxies/arcmin$^2$, which yields a $<0.6\%$ amplitude in $\delta_G$ for the mean blending
effect. This is suppressed by $\bar{n}_F/\bar{n}_G$ {\blue when we only consider the contribution of blending to the
cross-correlation to foreground galaxies,  $\bar{n}_F$. The actual density of our samples are much
lower because we impose restrictive photometric redshift cuts (as shown in Fig.~\ref{fig:dNdm-i}), which may also correlate with the blending effect and complicate predictions.
Despite this, } when including this blending effect 
we find that the blending contribution is negligible  in our case:
it is always much smaller than our
errorbars in the cross-correlation measurements (we show results for this in later figures, but they can not be distinguished by eye). 

In our analysis we  attempt to separate the effect of blending from other obscuration effects by repeating the analysis for a sample where we remove all potential blended objects. As mentioned above, 
this can only be done here because we  use space based data with higher resolution to identify objects that are merged into single objects in ground based observations. 
More generally, as detailed in Dawson et al. (2012), the ability to recognize a blend is a function of scale.
Adding fluxes of objects that are potential blends produces negligible differences in our results below.

\subsection{Separating Magnification and Obscuration}
\label{sec:sep}

Magnification and obscuration  produce different effects in  the galaxy counts (see Eq.~\ref{Eq:counts2}), and the galaxy magnitudes (see Eq.~\ref{Eq:mag2}).  We can therefore combine these observations to separate both effects.

\begin{figure}
\includegraphics[width=3.3in]{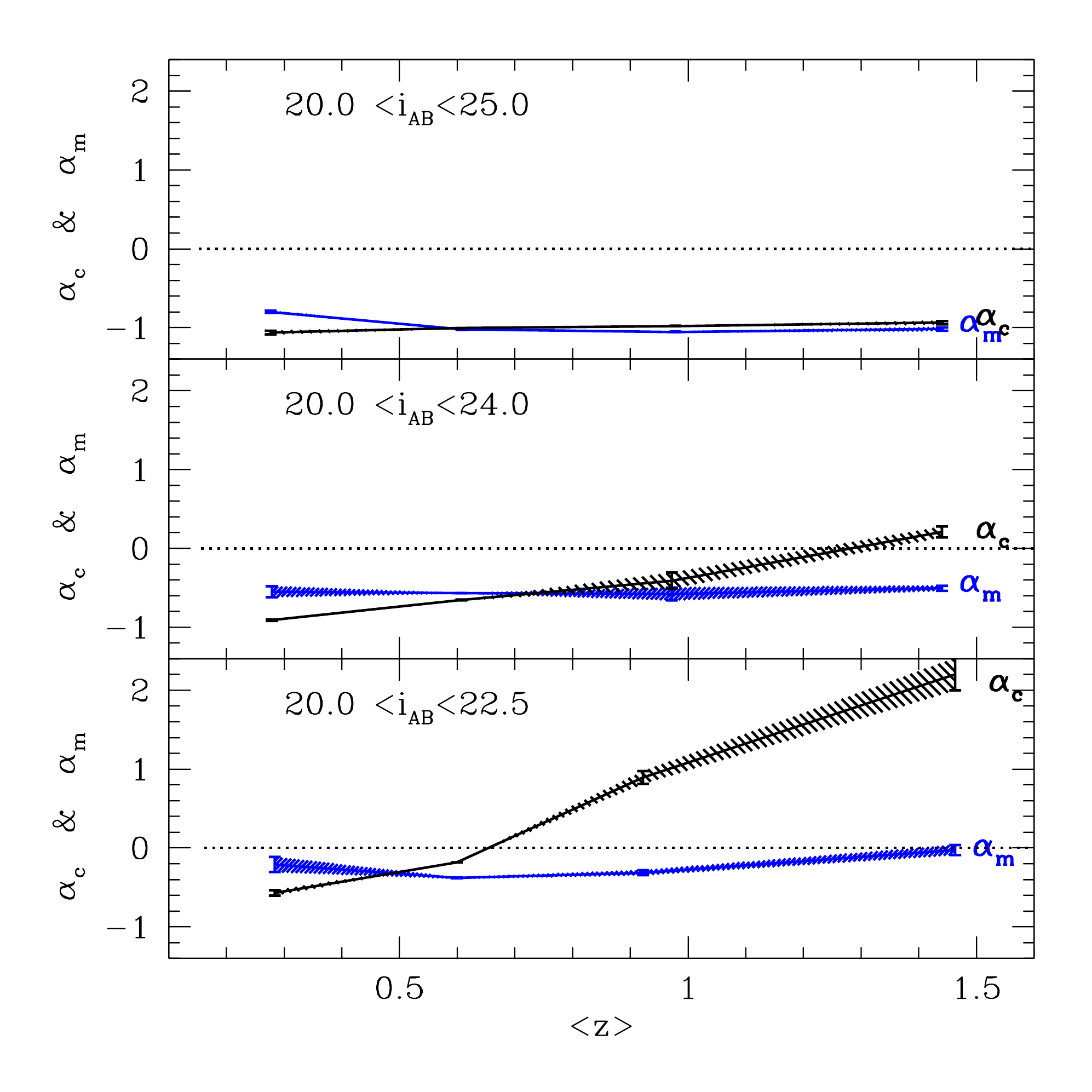}
\caption[]{The magnification amplitude is proportional to the 
number counts slope $\alpha_c$ (black, defined in Eq.~\ref{Eq:counts}) and mean magnitude slope $\alpha_\lambda$ (blue, defined in Eq.~\ref{Eq:mag}), which are shown here as a function of the mean redshift of the source sample. Top panel corresponds to the main samples used in this paper ($20.0<I_{auto}<25.0$) for which the slopes are nearly constant (after photo-$z$ selection, see Fig.~\ref{fig:dNdm-i}).  For comparison, middle and bottom panels show the results for brighter samples (which we will also used in the test of Fig.~\ref{fig:testcross})
 where the slopes can be closer to zero or positive.}
\label{fig:slope3}
\end{figure}

The corresponding proportionality constants $\alpha_c$ and
$\alpha_m$,  are defined in Eq.~\ref{Eq:counts} and \ref{Eq:mag}.
Fig.~\ref{fig:slope3} shows values of $\alpha_c$ and $\alpha_m$
estimated for different subsamples.
Errors in Fig.~\ref{fig:slope3}  correspond to the difference between measured numerical slopes using different magnitude increments ($dm_*=0.01$ compared to
$dm_*=0.02$).  
The top panel is the one relevant for our analysis, the other panels are included to illustrate how this could change for brighter samples. As mentioned above in Fig.~\ref{fig:dNdm-i}, the photo-$z$ selection (99\% C.L.) suppresses the  counts at the faint end of the sample and the slopes are close to $\alpha \simeq -1$. This maximizes the magnification signal and also makes the predictions insensitive to the uncertainties in the slope $\alpha$ estimation. The drawback is that we reduce the number of objects and bias the sample brighter than the parent flux limit sample.

We can now estimate the obscuration $A_\lambda^*$
in band $\lambda^*$ by combining $\Delta_\lambda^*$ and $\delta_G$:

\beq
A_\lambda^* = {2.5\over{\ln 10}}
\left[ {\alpha_c\over{\alpha_\lambda^*}}\Delta_\lambda^* -\delta_G  -B n_G^{2.15} \right]
\eeq
where $\lambda^*$ correspond to the band where the (flux limited)
sample is selected. For other bands $\lambda$:
\beq
A_\lambda = A_\lambda^* + \Delta_\lambda - {\alpha_\lambda\over{\alpha_\lambda^*}} \Delta_\lambda^* 
\label{eq:obs} 
\eeq

Magnification can then be found as:
\beq
\delta_\mu = {\Delta_\lambda^*\over{\alpha_\lambda^*}} 
+ A_\lambda^* {\ln 10\over{2.5}}
\label{eq:mag}
\eeq
In our analysis we will select samples with $\alpha_c \simeq \alpha_\lambda^* \simeq -1$, for which $\delta_\mu \simeq -\delta_G$ as obscuration only affects fluxes (i.e. $\Delta_\lambda^*$) and cancels out for counts.

Note that this estimator for $\delta_\mu$ only uses $\lambda^*$, the wavelength where galaxies are selected
(which in our case is the $I_{auto}-band$). The other bands can not be used here because we have used the information to determine the size of the obscuration correction. To have separate estimates of magnification from different bands we need to combine flux and counts estimates of galaxies
selected in the different bands. Here we only consider the $i-band$ selection because we are using the publicly available COSMOS 30-band catalog of Ilbert et al.~(2009),
which is only provided in the $i-band$ selection. But, we will show results for the obscuration in several bands to see how important this effect is as a function of wavelength. To be able to select in other bands we would need PSF corrected total magnitudes and a catalog that is uniform in that band.

\begin{figure*}
\includegraphics[width=6.in]{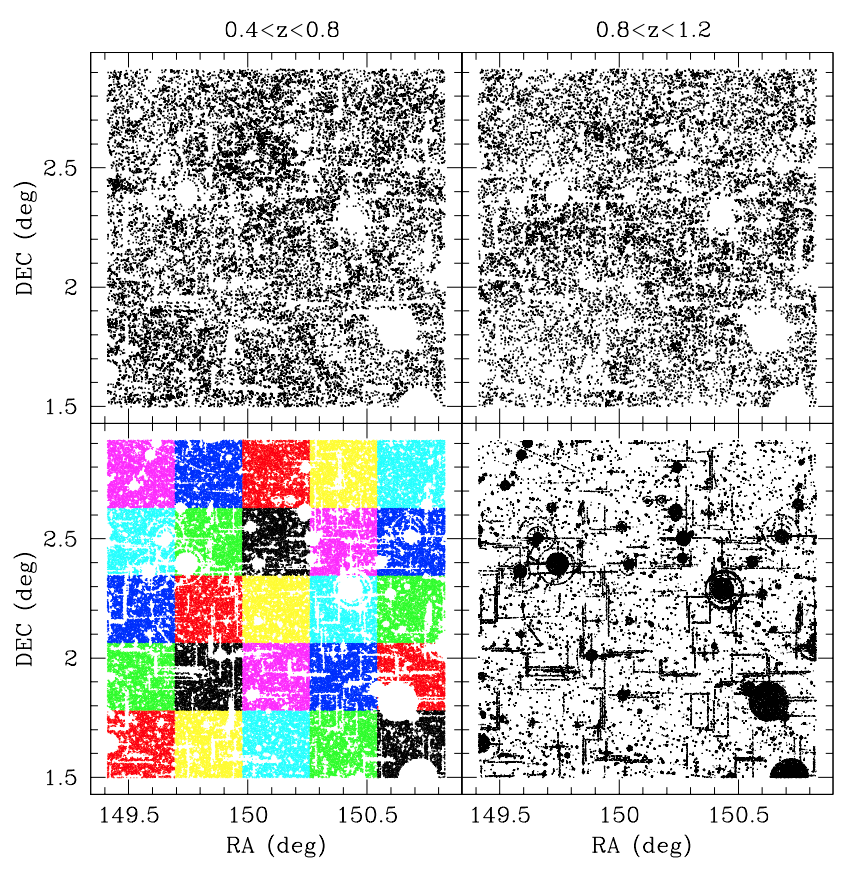}
\caption[]{Galaxy positions of the  $I_{auto}<25.0$ COSMOS samples used in this paper: 
lenses $0.4<z<0.8$ (top left) and sources $0.8<z<1.2$ (top right).
For flux measurements we use different bands: U, B, G, V, R, I, Z and K as given in the public version of the COSMOS catalog of Ilbert et al.~(2009).
Bottom left panel shows the random catalogs with
different colors for each of the 25 jack-knife regions. Bottom right panel
shows the mask.}
\label{fig:pij}
\end{figure*}

\subsection{Cross-Correlations}
\label{ssec:cross}

{\blue 
As mentioned previously, we can not estimate the effects of magnification directly from $\delta m$ in Eq.~\ref{eq:deltam}
or $\delta_G$ in Eq.~\ref{eq:deltaG}
for lens  galaxies because we do not know the intrinsic
fluxes or intrinsic counts before magnification or obscuration. 
Instead, we can estimate an average value for a population of lenses by comparing to the mean in a given galaxy selection:
$\Delta_\lambda$
in Eq.\ref{eq:delatL} or $\delta_G$ in Eq.\ref{Eq:counts2a}.

To estimate the population values $\Delta_\lambda$ and $\delta_G$  we  stack the background observables (i.e.sources, labeled as $j$) around the foreground galaxy (i.e. lenses, labeled $i$) positions:
\bea
<\delta_{G_i}\delta_{G_j} > (\theta)  &=& \frac{1}{ \sum_{i,j} 1} \sum_{i,j}  \delta_{G_i}\delta_{G_j} 
\nonumber \\
< \delta_{G_i} \Delta_{\lambda_j} > (\theta)  &=& \frac{1}{ \sum_{i,j} 1}  \sum_{i,j}  \delta_{G_i}  \Delta_{\lambda_j}
\label{eq:cross}
\eea
binned as a function of the $(i,j)$ pair separation $\theta$, the radius of an annulus $d(i,j) \in \theta \pm d\theta$.}
We estimate  $\delta_G = n/\left<n\right> -1$  using uniform random (positions) catalogs (excluding masked and boundary regions, as shown in the bottom right panel of Fig.~3) to define the mean density $\left<n\right>$ that is available around each annulus. We also use random catalogs to estimate the mean magnitude {\blue
$\bar{m}_\lambda \equiv <m_\lambda>$ in $\Delta_\lambda = m_\lambda - <m_\lambda>$ in each annulus}. We assign magnitudes randomly from the real catalog to the random position catalog and this results in a mean that is independent of position.

The dominant contribution to 
the fluctuations  $\Delta_\lambda$ and $\delta_G$ around each lens galaxy is not magnification,
but the intrinsic clustering of galaxies. We can cancel this contribution by averaging over the full
lens distribution,  where we stack the background galaxies for all foreground galaxies. This is equivalent to doing the counts-counts cross-correlation. We have validated the cross-correlation results presented here against two independent cross-correlation codes over the same samples.

To interpret the results we assume that two samples are well separated in
redshift space so that there is no intrinsic contribution to the cross-correlation
from overlapping galaxies. This will be tested in \S\ref{sec:photozVal}.

\section{DATA}

We use the public COSMOS 30-band photometric redshift catalog of Ilbert et al.~(2009) in the
2-deg$^2$ COSMOS field.
We limit our study to $19<I_{auto}<25$ for both source and  lensed samples, and to the 1.64 sq.~deg  region that overlaps with the HST-ACS catalog of Leauthaud et al.~(2007) to identify potential blended objects.
The results are robust to different sample selections which 
we have explored. We have not tried to optimize the sample selection to get better
results, but rather use the simplest case to focus on the blending and obscuration effects.
Fig.~\ref{fig:pij} shows the galaxy distribution in COSMOS catalog, the random catalogs, the jack-knife-regions 
 and the mask that we have used. We split COSMOS in 5x5 grid regions (shown as color in the random catalog on the bottom left panel of Fig.~\ref{fig:pij}) to use as jack-knife-regions for errorbar estimation (see Norberg et al.~2009). We use a conservative mask (bottom right panel in Fig.~\ref{fig:pij}) to avoid impact of background light from star halos on galaxy photometry.

For flux measurements we use different bands: U, B, G, V, R, I, Z and K as given in the public version of the COSMOS catalog of Ilbert et al.~(2009). For the i-band we use $I_{auto}$ which is a total magnitude corrected for PSF and provides a uniform selection (the parent sample was defined to a depth at least one magnitude deeper in $I_{auto}$-band ($I_{auto}<26$). For the other bands we use the fixed SUBARU aperture photometry provided in the catalog, which is optimal for photo-$z$ estimation, but non-optimal for magnification measurements. These are only used for the flux magnification and not for selection of the samples or estimating count fluctuations. The goal is to assess the importance of obscuration effects in different bands.

\subsection{Photo-z selection and validation}

\begin{figure}
\centering
\includegraphics[width=3.in]{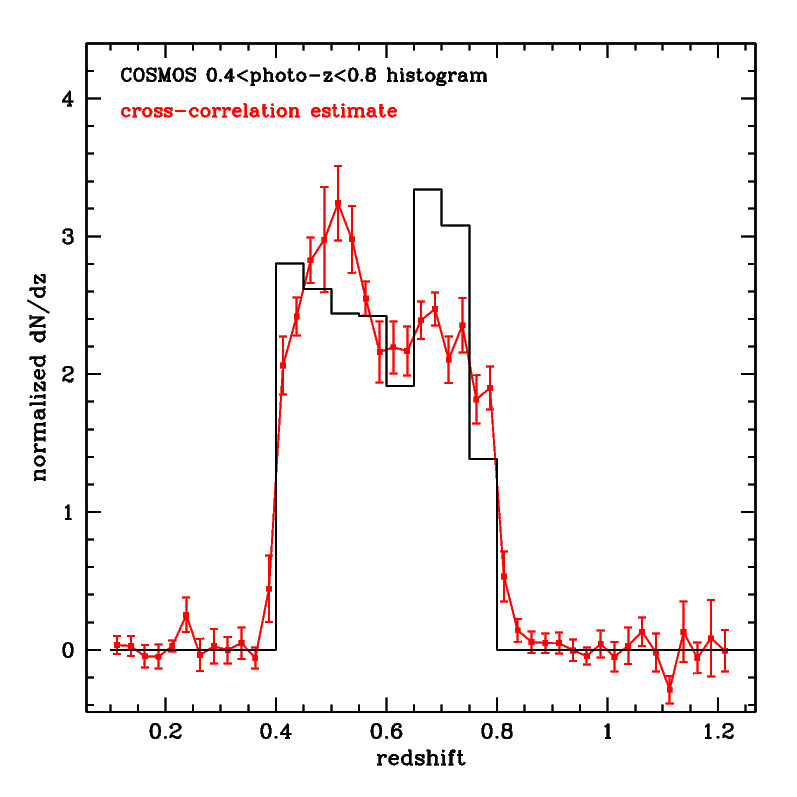}
\centering
\includegraphics[width=3.in]{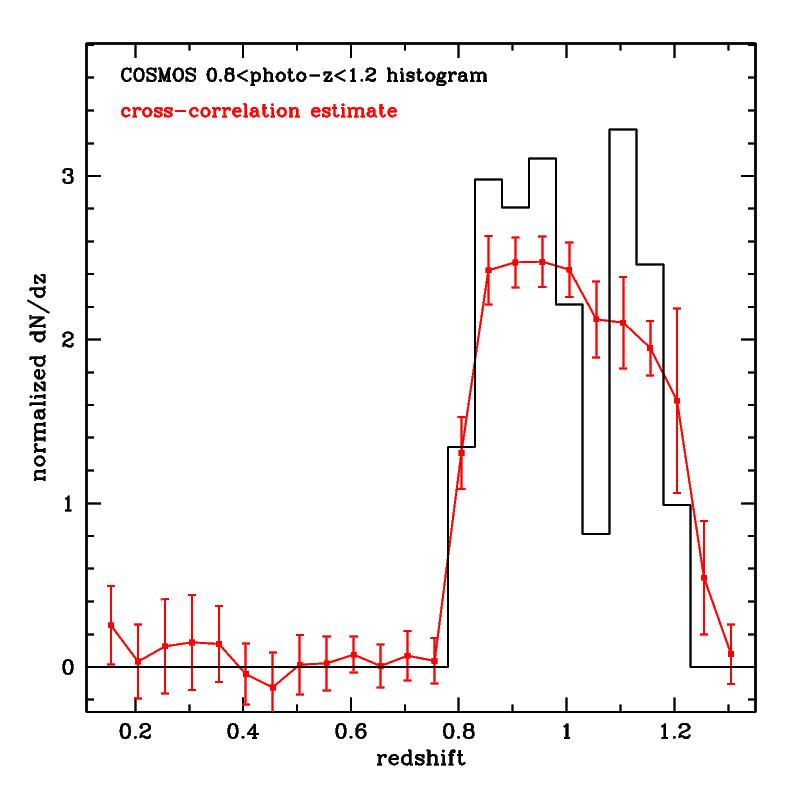}
\caption[]{Comparison of redshift distributions from best photo-$z$ values
(black histograms) and from  cross-correlation with a
spectroscopic sample from zCOSMOS (red lines and errorbars). Top panel corresponds to $0.4<z<0.8$ while bottom panel
has $0.8<z<1.2$. Cross-correlation estimates of the bin N(z) are consistent with those of the high quality photo-$z$ binning.}
\label{fig:nphotoz}
\end{figure}

We first explore the impact of photometric redshift errors on galaxy selection.  Ideal samples would contain no overlap in redshift between the lens and source samples, as this overlap could include a small but significant auto-correlation signal that could contaminate the lensing measurement.  In order to accomplish this sample separation we select only
those galaxies which have $99\%$ of their individual PDF inside the
chosen z-bin range.  Figure \ref{fig:nphotoz}  shows 
the density distribution $dN(z)/dz$ of best photo-$z$ values
normalized to the total number of galaxies, and shows that the conservative 99\% cut ensures that our lens and source bins are well separated. 
Table \ref{table:samples} shows statistics of these samples.

\begin{table}
\begin{tabular}{cccccc}
\hline
 $z$-bin & $\left<z\right>$ & $N_g$  & $<I_{auto}>$ & $\alpha_c/\alpha_m$  & $n_g$\\
$z$ range & mean & $\#$ & mag & & $\#$/arcm$^2$ \\
0.4 - 0.8 & 0.61 & 23532 & 22.48 &  & 5.0 \\
0.8 - 1.2 & 0.98 & 19581 & 23.10 &  -1.0/-1.0 &  4.1\\
stars  &  & 10088 & 21.98 & & 2.2\\
\hline
\end{tabular}
\caption{Background (sources) and foreground (lenses) samples used in this paper. Both samples have
 $19<I_{auto}<25.0$ and $99\%$ of each galaxy individual photo-$z$ pdf is within $z$ range. As a null test we also use a sample of stars in the foreground with same magnitude cuts. Stars are brighter and more sparse.}
\label{table:samples}
\end{table}

\label{sec:photozVal}

We validate the  $dN(z)/dz$ estimates using cross-correlation with spectroscopic redshifts, following the 
technique described in Schmidt et al.~(2013). We cross-correlate our 99\% photometric redshift cut samples with spectroscopically confirmed galaxies from the final release of the zCOSMOS bright sample (Lilly et al.~2007). The validation is shown as lines with error-bars in
Fig.~\ref{fig:nphotoz}. The top panel corresponds to 
$0.4<z<0.8$, while the bottom panel has $0.8<z<1.2$.
By comparing the two panels we can see that there is
little overlap between the two samples.
As expected by our selection, the overlapping tails 
between bins are smaller than 1\% which have negligible 
effects on the cross-correlation measurements, given our
error-bars and the amplitude of the signal.
{\blue Results are relatively insensitive to specific details of the redshift bin choice. We choose a z-bin width or $\Delta z =0.4$ in order to define a large enough sample to measure a significant correlation signal, and start at $z=0.4$ in order to define a sample with a substantial cosmological volume.
}

\subsection{Cross-correlation validation}

Here we test if our cross-correlation measurements are consistent with the weak lensing magnification signal.

\begin{figure}
\center
\includegraphics[width=2.7in]{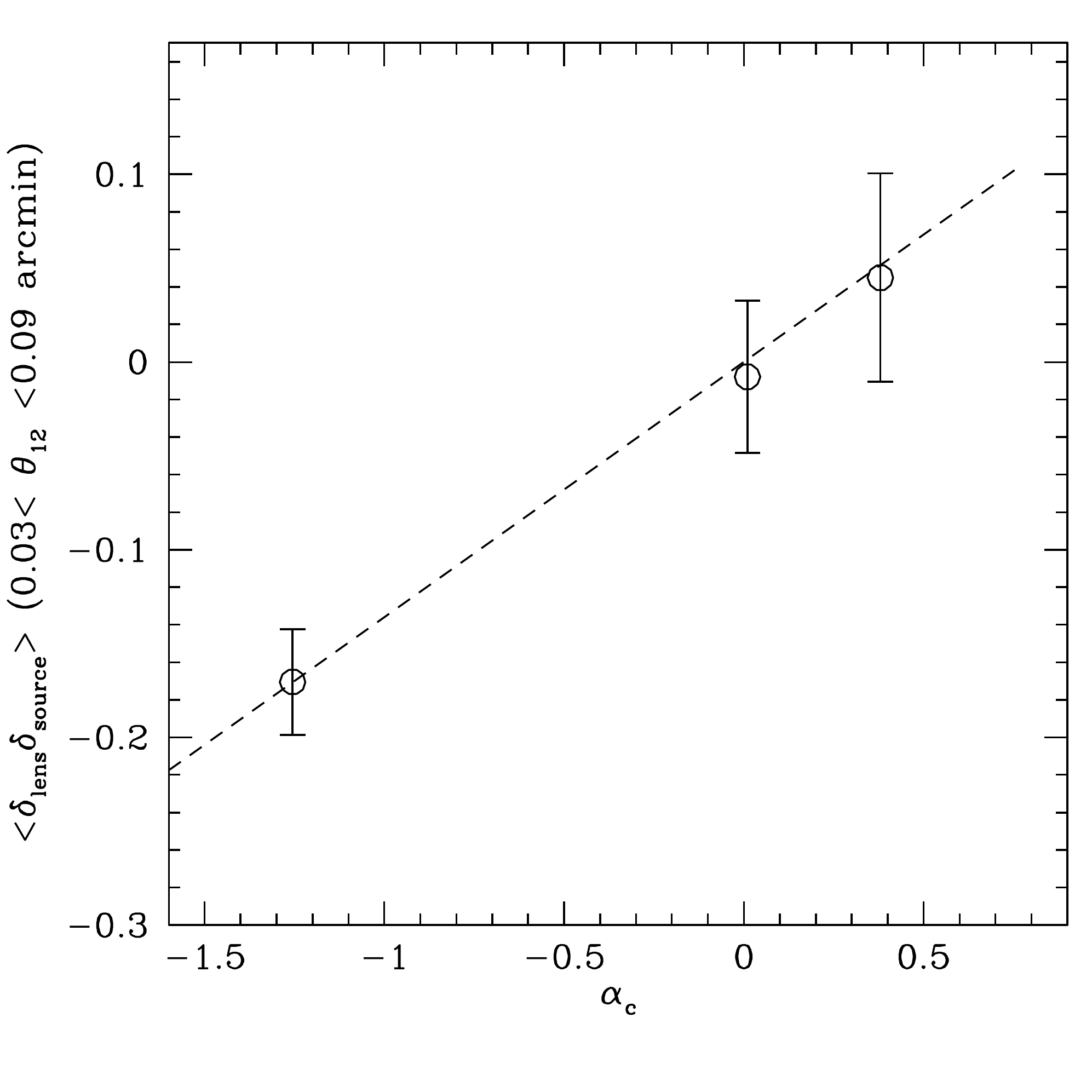}
\caption[]{Amplitude of counts cross-correlation  at small scales ($0.03<\theta_{12}<0.09$ arcmin)
between a lens sample at $0.4<z<0.8$ and a source sample at  $0.8<z<1.2$ 
as a function of $\alpha_c$. Each point corresponds to one in three  ($j=1,2, 3$) different magnitude cuts ($<I_{auto}(j)>=$23.6, 23.0 and 22.7)  which result in $\alpha_c(j)=$-1.26, 0.01 and  0.38. The dashed line is the prediction in Eq.\ref{Eq:counts} based on the first measurement. The scaling with $\alpha_c$ agrees well
with that expected for weak lensing magnification. }
\label{fig:testcross}
\end{figure}

 We select three ($j=1,2, 3$) different source galaxies (all within $0.8<z<1.2$)  with $22.5<I_{auto}(1)<25.0$, $20<I_{auto}(2)<24.4$ and  $20.0<I_{auto}(3)<24.0$ which have slopes:  $\alpha_c=$-1.26, 0.01 and  0.38 and mean magnitudes 
$<I_{auto}>=$23.6, 23.0 and 22.7. We test if the cross-correlation with the same lens sample  ($0.4<z<0.8$ as in Table \ref{table:samples})
scales linearly with the number count slope $\alpha_c$, as expected from Eq.~\ref{Eq:counts}. This test is shown in  Fig.~\ref{fig:testcross}, which displays the amplitude of the cross-correlation in the first angular bin (where S/N is largest) as a function of $\alpha_c$.  The dashed line shows the prediction based on the first point: $ \delta_G(j) =   \alpha_c(j)/\alpha_c(1)  ~\delta_G(1)$ (with $\alpha_c(1)= -1.26$). There is a very good agreement, which indicates that the cross-correlation signal is consistent with the weak lensing magnification effect. Note in particular  the null test (middle point) where we find no cross-correlation for $\alpha_c \simeq 0$, as expected. Brighter fluxes give larger slopes (see Fig.~\ref{fig:slope3}), but they also include fewer galaxies and larger cross-correlation errors (because of the small area of COSMOS), so this regime has lower S/N and is more difficult to test with COSMOS.


\section{Results}

In this section we apply the methodology presented in section 2 to the COSMOS samples defined in section 3. The goal is to combine the two cross-correlation signals from fluxes and counts to estimate the obscuration effects and recover a cross-correlation signal that is consistent with the magnification effect. We then estimate the mass profiles of the lenses and compare the results with simulations. We end with a comparison of the corresponding profiles obtained using tangential shear and flux-size magnification for the same lenses.

\subsection{Raw cross-correlations}

\begin{figure}
\center
\includegraphics[width=3.1in]{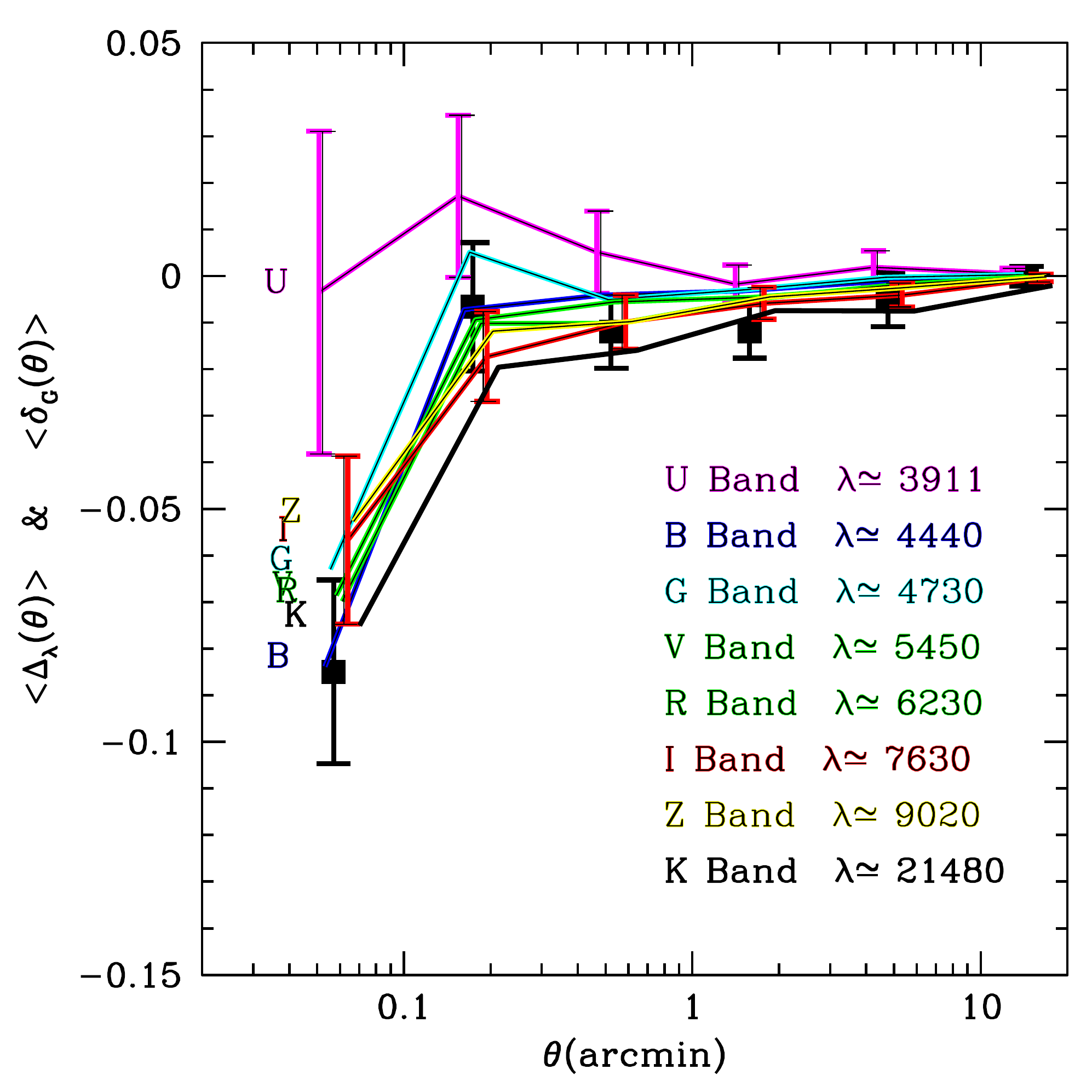}
\caption[]{Source counts fluctuations $\left<\delta_G\right> $ (filled squares) and magnitude differences $\left<\Delta_\lambda\right>$
in different bands (color lines) for galaxy sources ($0.8<z<1.2$)
around galaxy lenses ($0.4<z<0.8$) in circular annulus of mean radius $\theta$. For clarity we only show errors for counts and for K, I, and U bands (black, red, and purple lines). These are the raw cross-correlations
without any correction for obscuration effects or scaling by slopes. Except for the U-band, all estimates seem to be consistent with each other.}
\label{fig:raw}
\end{figure}

We first present here the raw cross-correlation results without any corrections. 
The angular cross-correlation of density fluctuations $\left<\delta_G\right>$ and magnitude fluctuations  $\left<\Delta_\lambda\right>$  in several bands $\lambda$, as detailed in Section~\S\ref{ssec:cross}.
We use six angular bins, which are uniform in log space in the range: $0.03 <\theta<30$ arcmin, but results are similar, within errors, for different bins.
Errors are estimated using the jack-knife samples, as explained in section 3. 
Fig.~\ref{fig:raw} compares the 
raw source count fluctuations
and magnitude differences of sources around lenses, i.e.  
$\left<\delta_G\right>$ and $\left<\Delta_\lambda\right>$ in \S\ref{ssec:cross},
using different bands: U, B, G, V, R, I, Z and K for $\lambda$.

Because in our case $\alpha_c \simeq \alpha_\lambda \simeq -1$ we can already see here that there is a good agreement between the magnification signal from counts and flux for most of the bands. This is a good internal consistency check  and shows that we can use different bands to estimate magnification.
The U-band shows the biggest outlier, indicating that is more affected by systematic effects (see below).

As a null test, we have also cross correlated  the stellar counts and magnitude fluctuations around the same lenses.\footnote{Note that we can not use a random catalog for a null test because we have already used a random catalog in our estimator (see \S\ref{ssec:cross}) and the cross-correlation should be zero by construction. We have indeed done this test for validation and find cross-correlation consistent with zero within shot-noise.}
In this case the errors are only slightly larger because there are only $N=10088$ stars in the sample (with same magnitude cuts as for galaxies) with a mean magnitude of $\left<I_{auto}\right>=21.98$ compared to  $\left<I_{auto}\right>=23.10$ in the galaxy sources.  Within errors there is no signal in the stars counts or fluxes, as expected (see dashed line in Fig.~\ref{fig:obs2}). This already indicates that obscuration (or merging) errors are not large, as compared to errorbars, at least for the brighter stellar sample.

\subsection{Effects of obscuration and blending}

\begin{figure}
\center
\centering \includegraphics[width=3.in]{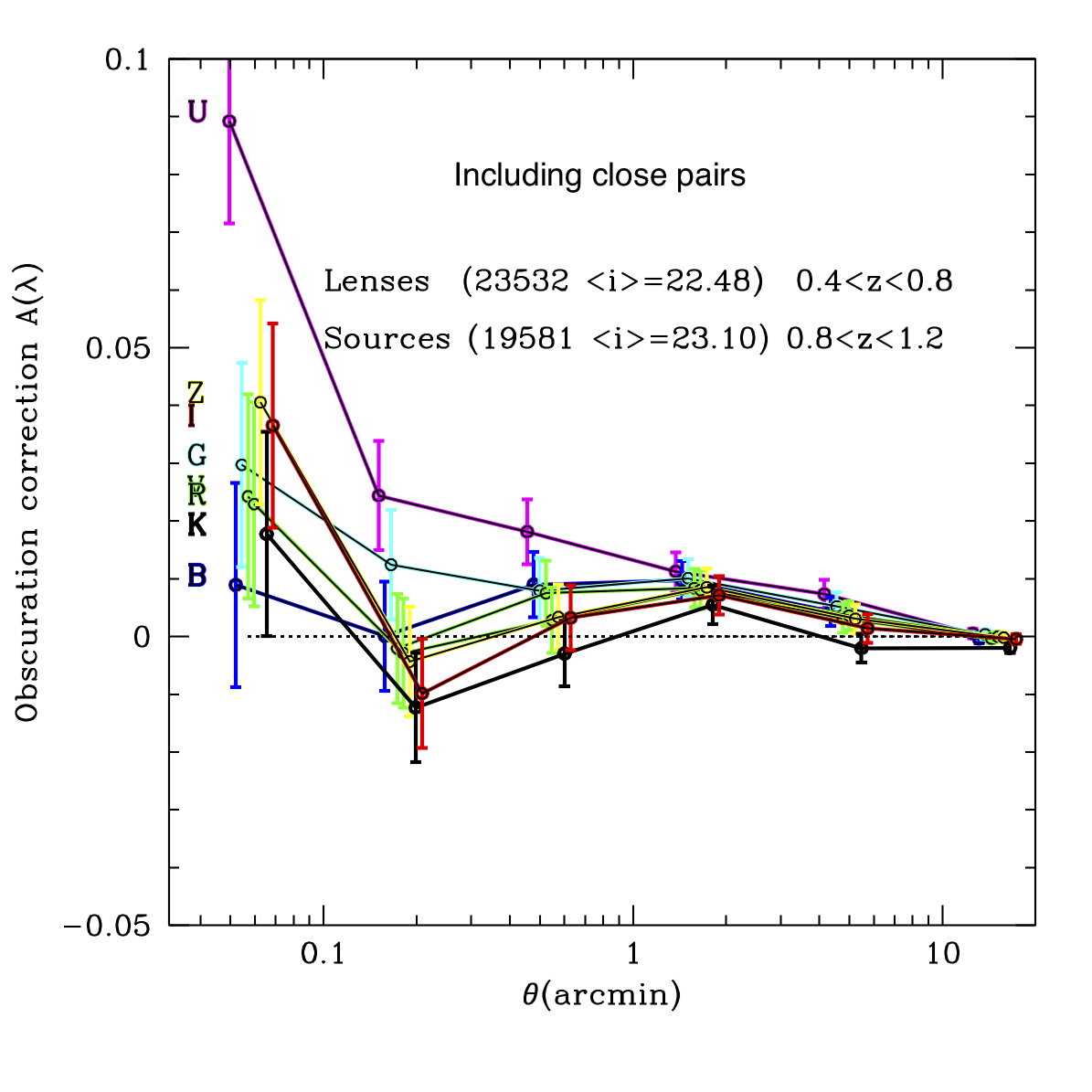}
\vskip -0.8in
\centering \includegraphics[width=3.in]{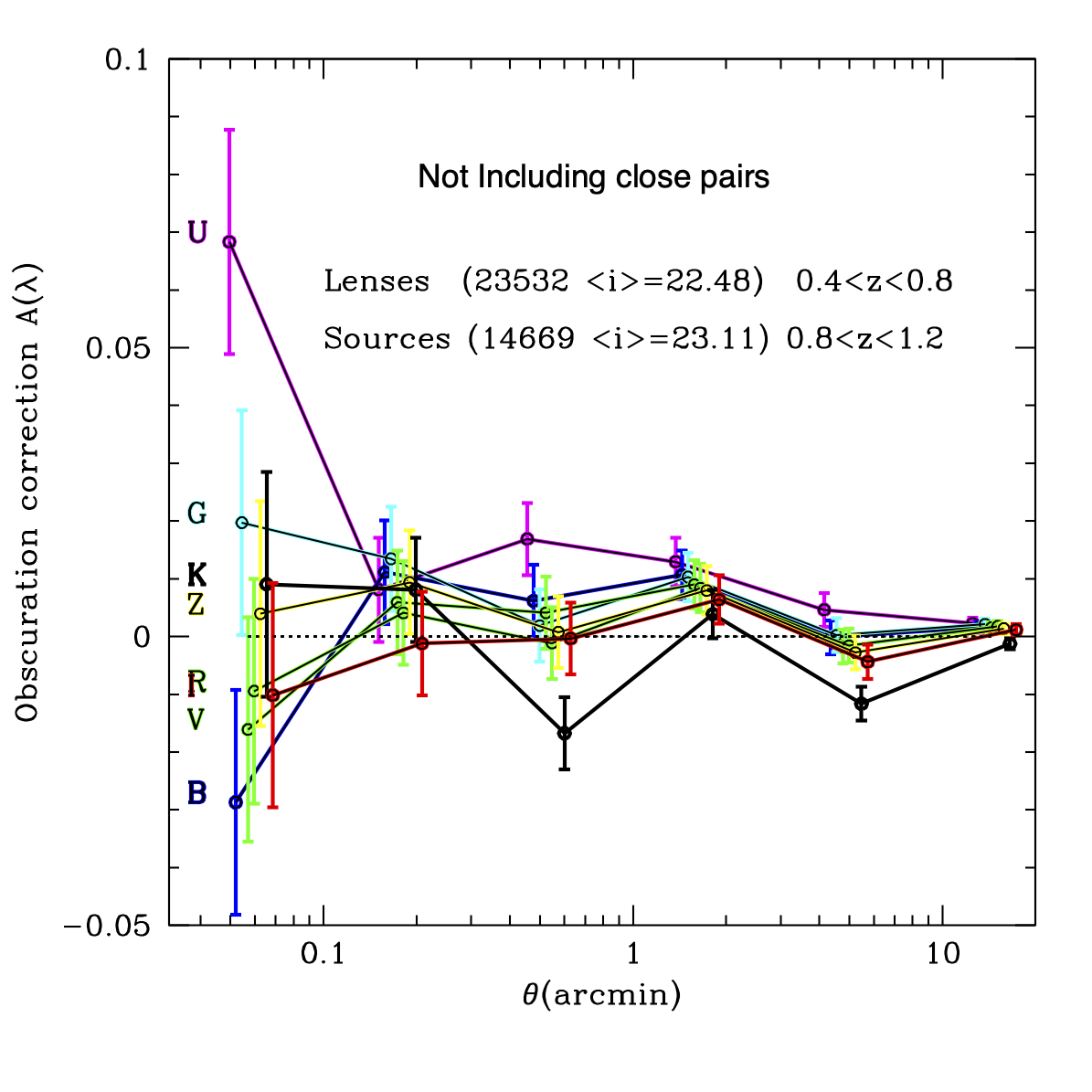}
\caption[]{Obscuration correction (from  
Eq.~\ref{eq:obs})
for different bands $\lambda$ when galaxies are selected in the I-band.
Color coding is as in Fig.~\ref{fig:raw}. The top panel include all galaxies, while the bottom panel exclude pairs that are too close (potential blends). Blending effects are important (up to 0.04 mag obscuration) at angular scales smaller than a few arcsec, but errors are too large to characterise this effect 
 with confidence as a function of $\lambda$.}
\label{fig:obs}
\end{figure}

From the above raw cross-correlations we can use the formalism presented in Section 2 to estimate 
the obscuration effect $A(\Lambda)$ from comparing the average flux to the average counts around lenses.
We estimate the obscuration for each jack-knife region and use it to estimate the errors as detailed in section 3. 
The top panel of Fig.~\ref{fig:obs} shows the  obscuration  correction
$A(\lambda)$ using Eq.~\ref{eq:obs}, which combines the effects of counts and magnitudes to separate obscuration from magnification.  {\blue Recall that the values of $A(\lambda)$ have units of magnitudes.
We can see that these corrections are small (less than 0.1 mag) but significant and need to be taken into account for accurate magnification estimates.}

The general trend shown in Fig.~\ref{fig:obs} is that obscuration is consistent with extinction
in that is larger for bluer bands and close to  zero for 
reddest bands. But there are exceptions to this (e.~g.~the B-band)  so that we can not exclude that some of these trends
are due to other systematics.

The bottom panel of Fig.~\ref{fig:obs} shows the same obscuration correction when we remove from the sample objects that are potential blends (galaxies in close pairs, separated by $\theta <1$ arcsec according to ACS-HST measurements, see \S2.4). The number of pairs is slightly reduced, which increases the uncertainties, however, the size of the errorbars does not increase much. 
In general, we can see that when using all galaxies, there is obscuration for all bands on the smallest scales, while this tendency tends to disappear when we remove the potential blends (with the exception of the U-band). 
This indicates that blending is playing some role in the obscuration signal and can not be ignored, even for our sample which is relatively bright on average because of the photo-$z$ selection ($<I_{auto}> \sim 23.1$, see Table 1).

\begin{figure}
\centering \includegraphics[width=3.in]{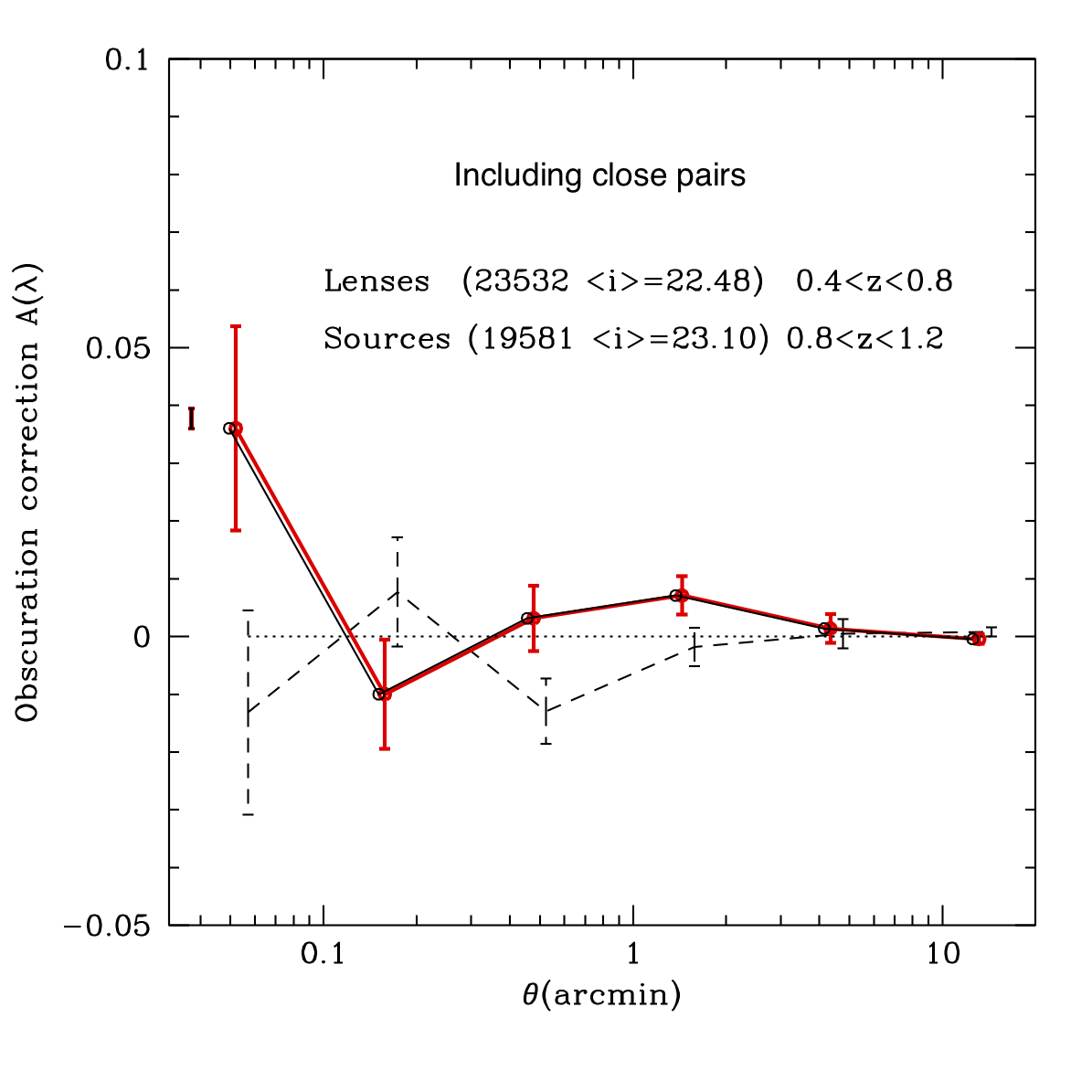}
\vskip -0.8in
\centering \includegraphics[width=3.in]{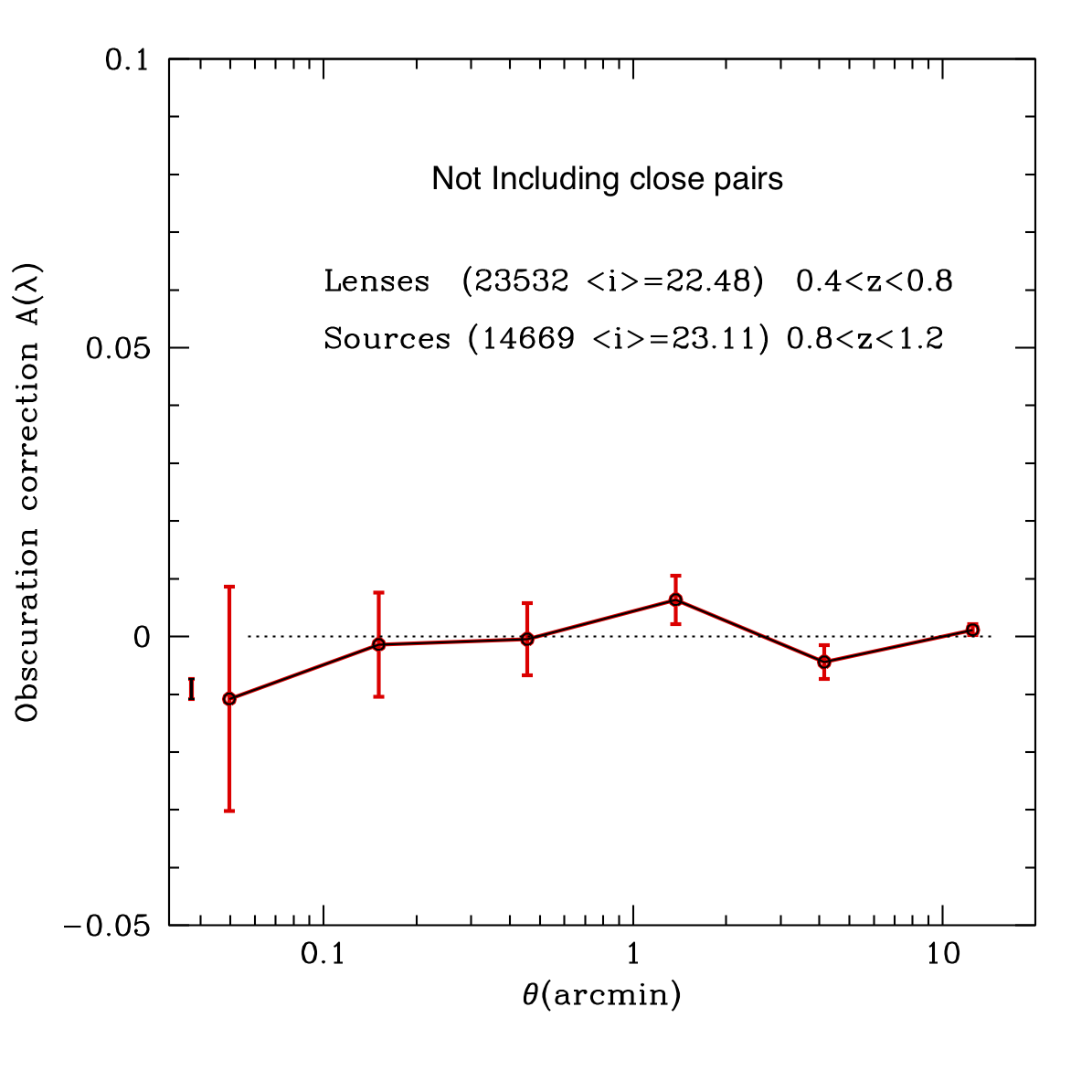}
\caption[]{Obscuration correction for the I-band ({\blue i.e. one of the curves in Fig.\ref{fig:obs}}).
The top panel includes all galaxies, while the bottom panel excludes pairs that are too close (potential blends). The dashed line in the top panel shows the corresponding obscuration correction when using the foreground stars in the field instead of the galaxy sources. Obscuration correction is significant for galaxies, but it is negligible
after removing the potential blends (bottom panel).}
\label{fig:obs2}
\end{figure}

We can see this tendency more clearly when we focus in the I-band case in Fig.~\ref{fig:obs2}. 
The correction for the I-band, which is the one most relevant for our magnification estimation, is consistent with zero at large scales but is significantly different from zero at the smallest scales.  After removing the potential blends (bottom panel) there is no significant obscuration correction.

The black lines show the mean corrected values for all  galaxies,  while  the  red  lines (which  lies  just  on top)  shows  the  results  without  the  average blending subtraction in Eq.~13

The dashed line in the top panel of Fig.~\ref{fig:obs2} shows the corresponding obscuration correction when we use stars instead of galaxies to correlate around the same lenses. This test indicates that the obscuration correction is negligible for stars, which could be explained by the fact that the stellar sample is over one magnitude brighter than the galaxy sample and their cross-correlation is negligible, as mentioned in previous section. Note that we are masking the area around bright stars with data reduction flags which also reduces the potential contamination in stars. We also note that de-blending effects in data reduction apply differently for stars and galaxies.

\subsection{Magnification estimates}
\label{sec:magnification}
 
\begin{figure}
\centering \includegraphics[width=3.in]{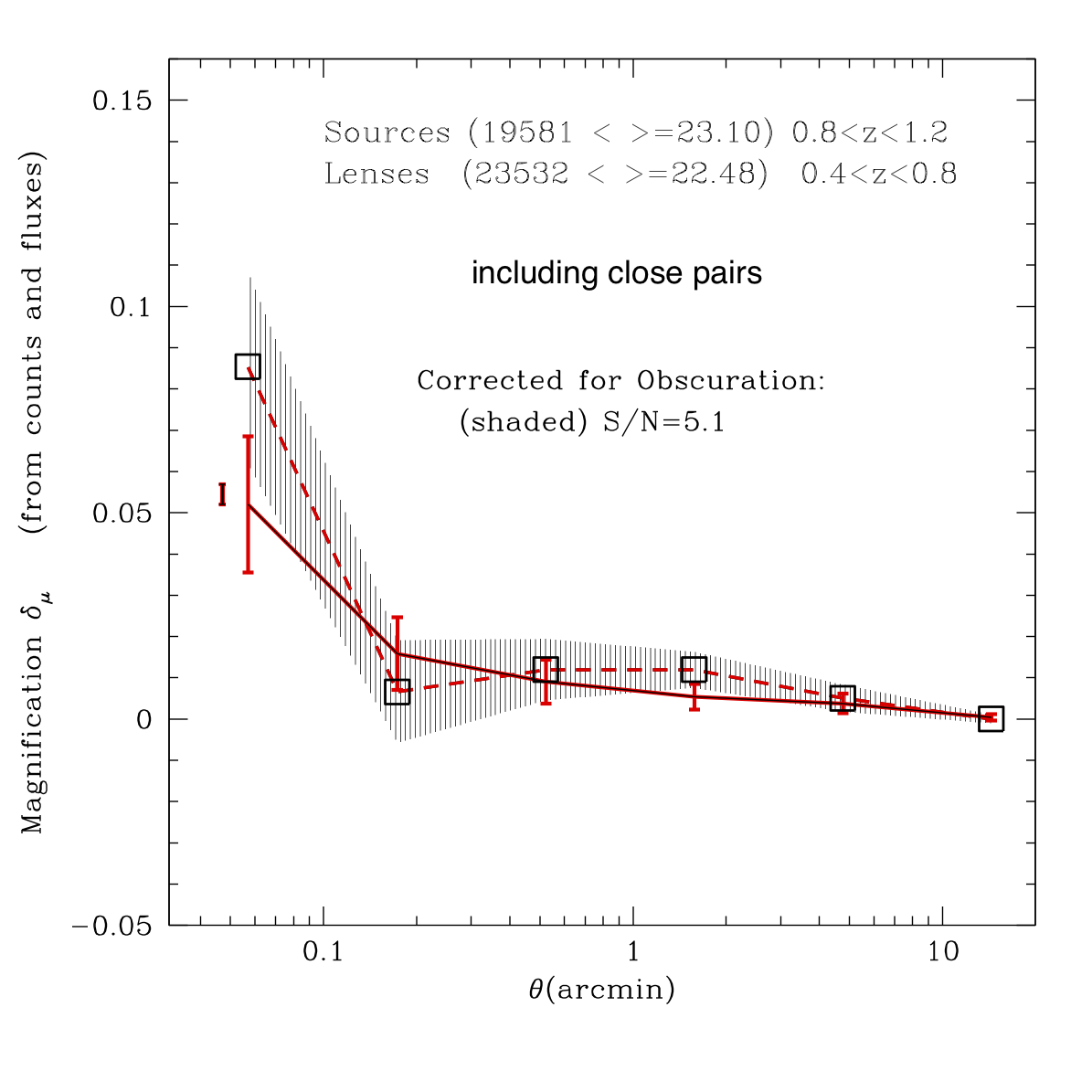}
\vskip -0.8in
\centering \includegraphics[width=3.in]{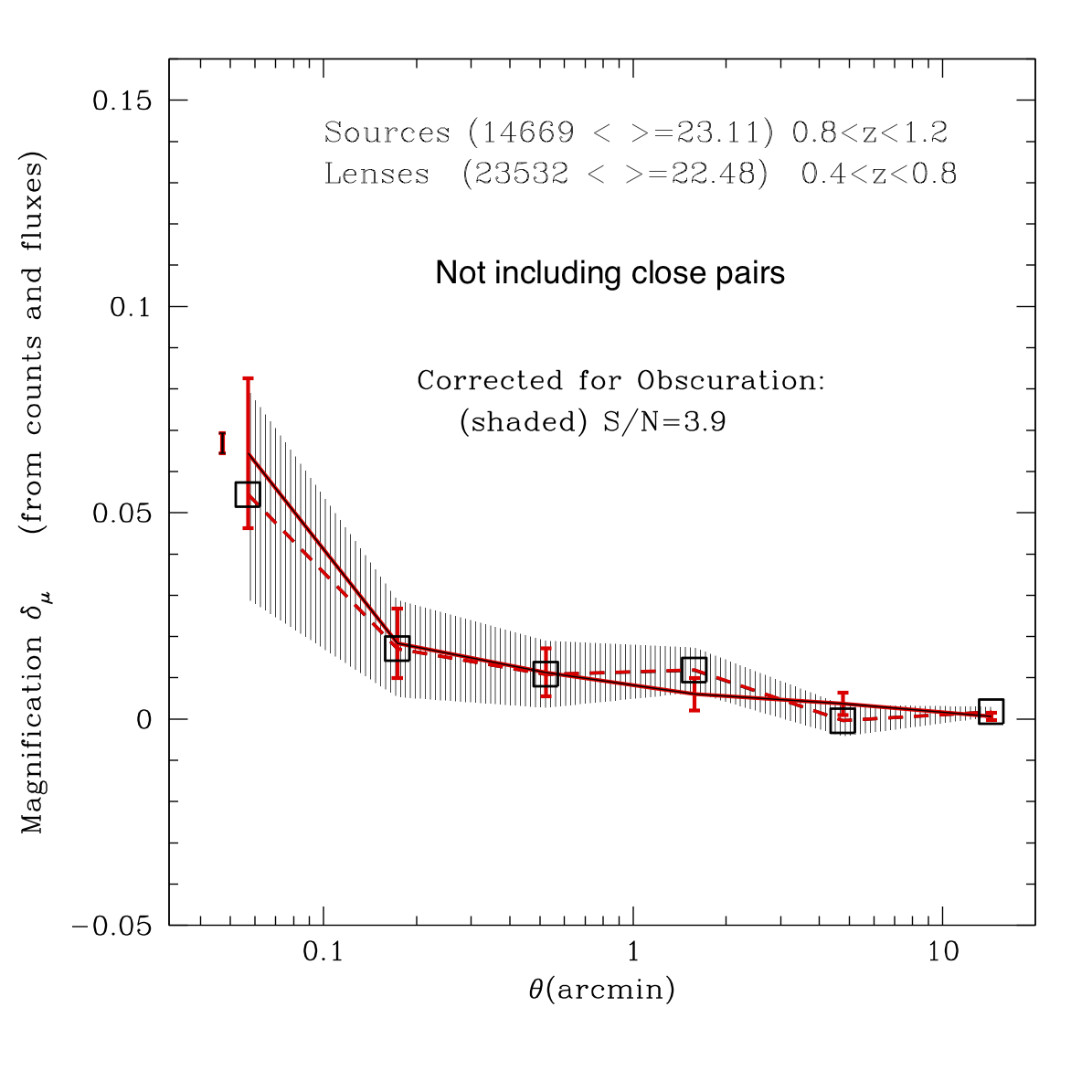}
\caption[]{Magnification with obscuration correction
(shaded region with dashed line) in the i-band as compared to the magnification estimate without
any correction from counts (squares) 
and from magnitudes (continuous lines).
The top panel includes all objects. In the bottom panel we remove potential blends (close pairs in HST images).
{\blue 
Removal of the close pairs does not cause a statistically significant shift in the measurement.}
Small scales have the highest S/N, but obscuration is also largest at these scales, see Fig.\ref{fig:obs}-\ref{fig:obs2}.}
\label{fig:mag}
\end{figure}

After obtaining the obscuration correction in previous sub-section, we now use it to recover the magnification signal combining flux and counts in the i-band.
We estimate the magnification for each jack-knife region and use the variance of the 25  jack-knife regions to estimate the errors.
Fig.~\ref{fig:mag} shows the  magnification estimates
with obscuration correction in Eq.~\ref{eq:mag} (shaded region)
and without corrections (lines and points), i.e.~scaled by the
corresponding slopes $\alpha_c$ and $\alpha_\lambda$. 
The grey dashed line shows the mean corrected values for all galaxies, while the red dashed (which lies just on top of the grey line) shows the results without the average blending subtraction in Eq.~\ref{eq:dn}. As mentioned before this term is negligible for our sample. But this does not mean the blending (or associated data reduction effects)  can not cause obscuration effects, as indicated in Fig.~\ref{fig:obs}-\ref{fig:obs2}.

Magnification from I-band flux has higher signal-to-noise than from I-band counts, though both measurements
are consistent after we include the obscuration correction, which only degrades the S/N 
slightly with respect to the uncorrected flux estimate.

The bottom panel of Fig.~\ref{fig:mag} shows the same results when we remove pairs that are potential blends (i.e. galaxies that have pairs closer than $\theta <1$ arcsec according to the ACS-HST). Close pairs are only present in the source galaxies sample ($0.8<z<1.2$). There are no close pairs in the lens sample ($0.4<z<0.8$) as the corresponding physical scales
are smaller and the sample
contains brighter galaxies (see Table 1). The number of sources is reduced from 19581 to 14669. Removing potential blends reduces the magnification signal (and the $S/N$) by about $\sim 30\%$.  {\blue Removing the close pairs results in measurements that fall within the 1-sigma uncertainties, showing that the close pairs are not having an appreciable impact on the magnification measurement.  However, out of an abundance of caution we remove close pairs in our calculations to avoid any potential systematic effects.}  In next sub-section, we use it to estimate matter profiles and compare to other results.
 
 
 \subsection{Matter profiles}
 \label{sec:matter}

The gravitational lensing effect is  proportional to
the projected mass density $\Sigma$, which on small scales
is typically dominated by the surface density of a single dark matter
halo, which acts as a lens:

\beq
\delta_\mu(\theta)  \simeq 2 ~\kappa(\theta)  \simeq 2 ~{\Sigma(\theta) \over{\Sigma_c}}
\eeq
The  weight,  $\Sigma_c$,  depends on the critical density
$\rho_c$ and  the lensing geometry, but is independent of wavelength $\lambda$:
\beq
\Sigma_c = {c^2\over{4\pi G}} {\chi_j \over{\chi_i \chi_{ij}}}
= {2\over{3}} {\chi^2_H \chi_j\over{\chi_i \chi_{ij}}} \rho_c
\eeq
where $\chi_H \equiv c/H_0$ and $\chi_i$, $\chi_j$, $\chi_{ij}$ refers to the comoving
distance to the lenses, to the sources and between them.

\begin{figure}
\centering \includegraphics[width=3.4in]{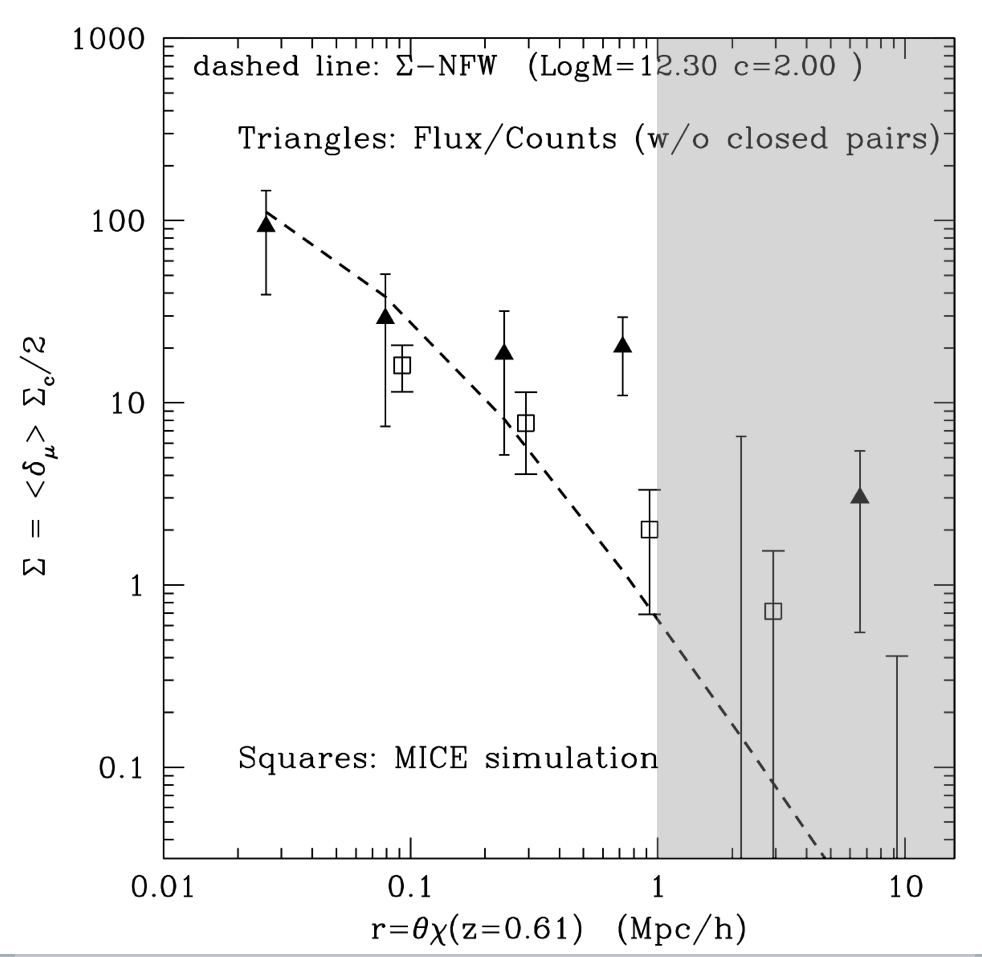}
\caption[]{Projected mass $\Sigma = \delta_\mu \Sigma_c/2$ from flux/counts magnification without close pairs (triangles) compared with results from the MICE simulations (squares). 
Dashed line shows the NFW 1-halo with $M_{200} \simeq 2 \times 10^{12} M_{\sun} h/pc^2$ and $c=2$. The mean simulation results show a clear (and significant) excess above the NFW profile on the largest scales, indicating the presence of a 2-halo term. The data is consistent with the 1-halo term on small scales ($<0.5$ Mpc/h) but are too noisy in the 2-halo scales (shaded region) because of the small size of the COSMOS field.}
\label {fig:sigma}
\end{figure}

Closed triangles in Fig.~\ref{fig:sigma} show the mass profile (in comoving coordinates $r=\chi_i \theta$) inferred from the magnification measurement 
$\delta_\mu$ (shown in bottom panel Fig.~\ref{fig:mag}) using the following relation:

\beq
\Sigma(r) = \delta_\mu(\theta)~ \Sigma_c/2
\eeq

We estimate the  mass profile for each jack-knife region and use the variance of the 25 jack-knife regions to estimate the errors. 
Fitting a power law to the triangles in Fig.~\ref{fig:sigma2}
we find a mean amplitude of dark matter projected mass  $\Sigma= 25 \pm 6 M_{\sun}h/pc^2$ 
at comoving radius of 0.1 Mpc/h.
 The dashed line shows the 1-halo NFW  profile with concentration $c=2$ that best matches this profile. We fit the first 3 points because the NFW is too steep to fit the measured profiles and we only expect the 1-halo term to dominate in the smallest scales. On larger scales ($>0.3Mpc/h$) the magnification measurements show marginal evidence for an up-turn (above the 1-halo term)  which could be related to the 2-halo term. However, note that the 3 last points in the profile are consistent with zero at a 2-sigma level of significance.

Our flux limited (photo-$z$ selected) lenses  have $<I_{auto}> \simeq 22.5$ at $<z> \simeq 0.61$ (see Table 1). This corresponds  to a k-corrected absolute magnitude of $M \simeq -21$ which matches a  NFW profile with mass of $M_{200} \simeq 2 \times 10^{12} M_{\sun} h/pc^2$ and $c=2$.

 \subsection{Comparison with simulations}
 
 We utilize simulations to validate our method, to test the error estimation and to compared the mass profile estimates with the LCDM model.
We use the MICE simulations (Fosalba et al. 2015a, Crocce et al. 2015, Fosalba et al. 2015b, Carretero et al. 2015, Hoffmann et al. 2015), which include the magnification effect both in magnitudes and positions (see Section  5 and Fig.14 in Fosalba et al. 2015b).\footnote{Simulations used here are available through the COSMOHub (Carretero et al. 2017) web page: \url{https://cosmohub.pic.es/home}}
In the simulations we use true cosmological redshifts instead of photometric redshifts, as the photometric redshift effects is already tested in Section 3.2.
We first limit the simulations to $19<I_{auto}<25$.
To include the same galaxy selection in  the simulation as in the samples of Table \ref{table:samples}, we draw objects randomly from the simulation to have the same $I_{auto}$ magnitude distribution in steps of $\Delta z =0.01$ and $\Delta I_{auto} =0.01$.

Fig.~\ref{fig:sigma}  shows as squares the results in the simulation.
 Here we show the $\Sigma$ measurements from the galaxy-magnitude cross-correlations for galaxies selected in a similar way as in Table \ref{table:samples}. Mass profiles in the simulation are only reliable for scales larger that $\simeq 0.1$ Mpc/h because of resolution effects (particle mass is $\simeq 3 \times10^{10} M_{\sun}/h$ and $l_{soft}= 50$ kpc/h).
 
 Error bars on the the square points in Fig.~\ref{fig:sigma} are derived from a set of 50 separate COSMOS-sized regions (of $\simeq 1.7$ deg$^2$ and using the same mask as used in this paper, see Fig.~\ref{fig:pij}). 
  We can see that the simulation errors are comparable to the jack-knife errors in the COSMOS galaxies. We do not expect these errors to be identical because jack-knife errors are only approximate and the simulations are not realistic in the obscuration effects, photometric errors or the photo-$z$ performance. 
  
  The MICE simulations follow a similar NFW profile as data at small scales. But note the systematic differences above $\simeq 1$ Mpc/h due to the 2-halo term. Even when the simulation errors shown are large, because they correspond to the COSMOS area, the mean simulated values have a much smaller error, as they come from 50 times larger area. Deviations from NFW seem larger in the real data than in the simulations, but as mentioned above, these differences are not significant at the 2-sigma level in the data. Moreover, the sampling variance errors in the simulation indicate that the COSMOS field is too small to provide a detection above $\simeq 1$ Mpc/h scales.

\subsection{Comparison with tangential shear}

\begin{figure}
\centering \includegraphics[width=3.4in]{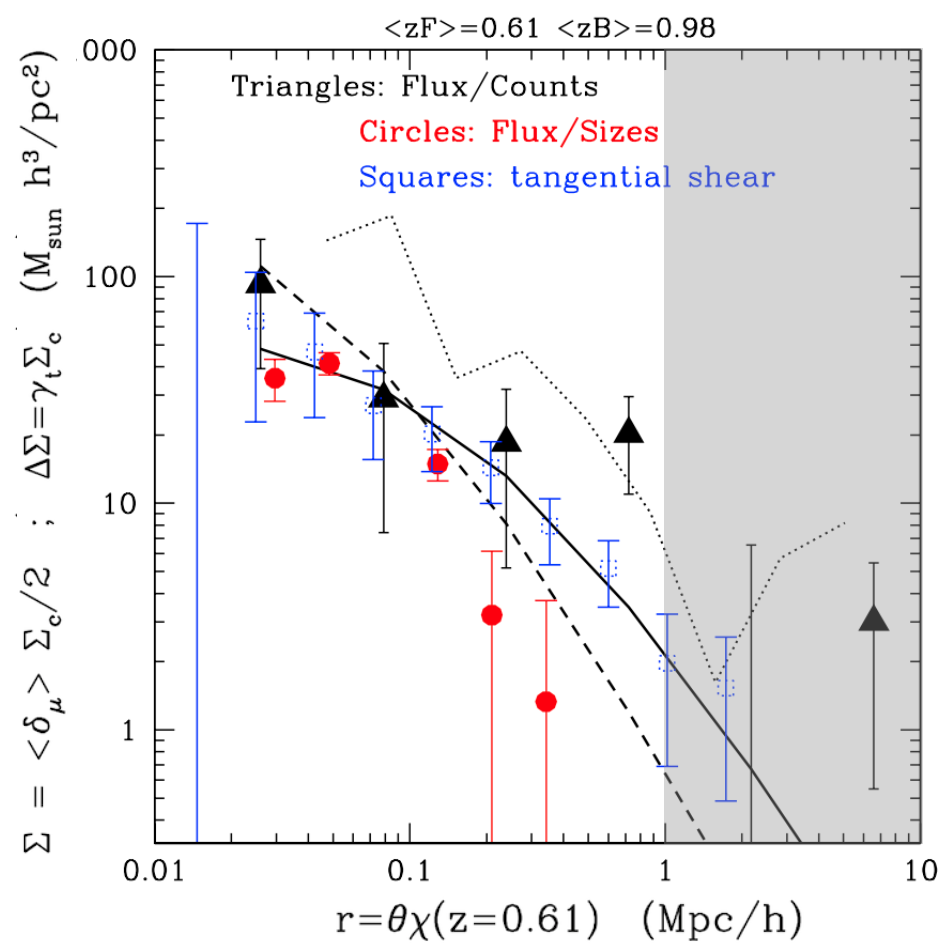}
\caption[]{
Projected mass $\Sigma = \delta_\mu \Sigma_c/2$ from magnification in flux/counts  (black triangles) compared to that from flux/sizes (red circles) and $\Delta\Sigma$ from tangential shear (blue squares with errorbars), all over the same galaxy lenses (first raw in Table \ref{table:samples}). The different methods agree well with each other and with 
the NFW predictions for $\Sigma$ (dashed line) and for $\Delta\Sigma$ 
(continuous line) with $M_{200} \simeq 2 \times 10^{12} M_{\sun} h/pc^2$ and $c=2$. 
Shaded  region correspond to scales that are too noisy for the area of the COSMOS field. 
The dotted line shows $\Sigma$ from Flux/sizes in galaxy groups (Schmidt et al. 2012). The group profile has similar shape to that of individual galaxies, but a factor of  4-8 larger amplitude for the density profiles, as expected.}
\label {fig:sigma2}
\end{figure}

With the results from the previous sections, we can now ask the question: are mass profiles estimated from magnification consistent with those derived from shear?
We measure differential mass profiles, $\Delta \Sigma$,  from tangential shear $\gamma_t$ around the same lenses  (see Schmidt et al. 2012):

\beq
\Delta \Sigma(\theta) \equiv \Sigma (<\theta)- \Sigma (\theta) =  ~\gamma_t(\theta) ~ \Sigma_c
\label{eq:dsigma}
\eeq
We then use the shear measurements of Leauthaud et al.~2007 to estimate tangential shear average over the same lenses (from sources in Schmidt et al. 2012). The result is shown in Fig.~\ref{fig:sigma2}. 
Profiles are shown as a function of comoving radius: $r=\theta \chi$ Mpc/h, where $\chi$ is the comoving distance to $z=0.61$, the mean redshift of the lenses, and $\theta$ is the measured angular separation.
To compare both profiles we show $\Sigma$ (dashed line) and $\Delta \Sigma$ (continuous line) for the same NFW profile (with $c=2$ and  $M_{200} \simeq 2 \times 10^{12} M_{\sun} h/pc^2$). We see how both measurements give similar profiles for the smallest comoving scales ($r<0.3 Mpc/h$). Our measurements from flux/counts (triangles) show a flatter profile, possibly due to the 2-halo term (see Fig.~\ref{fig:sigma}), while the tangential shear measurements seem to follow the 1-halo term. Note that $\Delta \Sigma$ in Eq.\ref{eq:dsigma} is in general less sensitive to  the 2-halo term than $\Sigma$ because the former is cumulative. However, the systematics  for $\Delta \Sigma$  and $\Sigma$
are different.

Our results for the 2-halo term (shaded region Fig.~\ref{fig:sigma2}) 
in are not conclusive.
Recall that the NFW profile shown was fitted to the three smallest scale points in $\Sigma$, which could bias this interpretation. The shear values could also be subject to some uncorrected shape biases. But note that magnification measurements have low significance on scales larger than $\simeq 1$ Mpc/h  (see Fig.\ref{fig:sigma}) and we need larger areas to be able to make a more detailed comparison on the 1-halo to 2-halo term transition.

\subsection{Comparison with flux/size and group profiles}

{\blue 
In addition to changing the flux of galaxies, magnification ($\mu$ in Eq.\ref{eq:mu}) also changes their apparent size (e.g. Bartelmann \& Narayan 1995, Bartelmann \& Schneider 2001). In this work, we have, to this point, not used measured galaxy sizes, but this information is available in the parent COSMOS catalog. 
We estimate mass profiles for the same lens sample (first row in Table \ref{table:samples})  with the method presented by
Schmidt et al. (2012) which combines flux and size magnification. This method has its own potential systematics which are discussed in Schmidt et al. (2012) and are not addressed here.} 

Is the flux-counts magnification signal consistent with the flux-size magnification signal?
Red circles  in Fig.~\ref{fig:sigma2} show the mass profiles estimated with flux-size measurements using the method presented in  Schmidt et al.~(2012),
over the same sample of lenses used in this paper. There is good agreement with the profiles from our flux-counts
magnification estimates on the smallest scales. On the largest scales, measurements have low significance and the flux-size measurements don't provide a detection. At 2-sigma level both results are consistent. We also plot as dotted line the flux-size measurements for galaxy groups in  Schmidt et al.~(2012). As expected the group estimates are larger:  $\Sigma \simeq 100-200  M_{\sun}h/pc^2$ at 
 comoving radius of 0.1 Mpc/h, four-eight times larger than for flux limited galaxies  ($<M_i> \simeq -21$ at $z \simeq 0.6$).
 
\section{Conclusions \& Discussion}

We have studied magnification with the COSMOS photo-$z$ Survey to $I_{auto}<25.0$ both with galaxy densities
and fluxes (magnitudes). We take advantage of the good photo-$z$ characterization to split foreground
($0.4<z<0.8$) and background ($0.8<z<1.2)$) galaxies into contiguous redshift bins  using photo-$z$ selection with 99\% CL. We estimate the redshift distribution $N(z)$ from
cross-correlation with zCOSMOS spectroscopic sample and find good agreement with the 99\% photo-$z$
selection. 

We introduce a  formalism to account for systematic obscuration effects (i.e. not due to magnification)
in the fluxes by combining counts and magnitudes. We find that obscuration effects are typically smaller than $0.05\ mag$ at all scales when galaxies are selected in the COSMOS $I_{auto}$ band,
but they need to be taken into account given the errorbars.
The general trend of obscuration is consistent with extinction
in that is larger for the U band and consistent with zero for the
reddest bands. On the smallest scales, where our results are more significant, this obscuration reduces significantly when we remove close pairs. This indicates that blending effects have an important impact ($\simeq 30\%$) in our magnification signal. 
 
We measure a significant ($S/N\simeq 3.9$) magnification signal that is consistent for counts and  magnitudes. 
 We  use these measurements to estimate the matter mass profiles of regular (flux limited)
($<I_{auto}> \simeq 22.48$) galaxies at redshift $z \simeq 0.61$ ($<M_i> \simeq -21$). We find
a mean amplitude of dark matter projected mass  $\Sigma= 25 \pm 6 M_{\sun}h/pc^2$  at comoving radius of 0.1 Mpc/h, which is consistent
with a NFW type profile ( $M_{200} \simeq 2 \times 10^{12} M_{\sun} h/pc^2$ and concentration of $c=2$ ) on scales $\simeq 30-300$ Kpc/h
and shows marginal evidence of  2-halo term at larger scales, up to 10 Mpc/h.
We compare these results with simulations and shear profiles (for the same lenses) which agree on small scales.

As shown in Fig.~\ref{fig:obs2} and Fig.~\ref{fig:mag} blending effects start to be important (up to a factor of $0.04~mag$ obscuration) at angular scales smaller than a few arcsec. But, due to the small area of the COSMOS field, errors are too large to characterize this effect with confidence. Extinction and other systematic obscuration effects can be as large as 0.1 mag but are typically smaller than 0.02 mag depending on the band (see Fig.~\ref{fig:obs}). These effects can quickly become more important for deeper samples, 
such as the deepest samples expected from the 
LSST survey (but not in the LSST Gold Sample).

Results in this paper are a proof of concept for magnification measurements using only galaxy counts and fluxes. We have not tried to optimize choices, such as sample selection, redshift binning or use of weights because the COSMOS field is too small to provide significant statistics.  Looking ahead, these methods are now being tested over  wider areas 
($\simeq 50-100$ deg2) with accurate photo-z, such as those in the PAU Survey (Eriksen et al 2019). As the statistical errors reduced, we will need better control of systematic effects and to explore ways to optimize our estimates.  In the near future we expect these methods to mature and
to be applied to the much deeper and wider (18,000 deg2) LSST samples. 
The COSMOS statistical errors presented in this paper should decrease by a factor of $\sqrt{10,000} \simeq 100$ for LSST, at the price of having a less accurate photo-z estimation.  The challenge then will be to have a better control of photometric and photo-z error uncertainties. Results with better photo-z precision from the smaller area surveys can then be used to help calibrate some of these uncertainties. 
Besides the methods explored in this paper to mitigate systematic effects, we should also use image simulations: embeding fake objects (with known properties) in real imaging to accurately characterize measurement biases (e.g. see Suchyta et al. 2016, Garcia-Fernandez et al. 2018). 

We conclude that magnification from counts and fluxes 
using photometric redshifts in flux limited galaxy samples 
has the potential to provide  accurate weak lensing measurements in future wide field surveys once we carefully take into account systematic effects, such as obscuration and blending.  These weak lensing measurements can be complementary (compared to cosmic shear and size magnification) for the deeper samples.

\section*{Acknowledgements}
We would like to thank Alex Alarcon and Laura Cabayol for help in validating the cross-correlation results with different clustering codes, and Imran Hasan and Erfan Nourbakhsh for assistance with some technical details of \textsc{TreeCorr}. We would also like to thank M.~Eriksen, for discussions on earlier stages of this project and Alexis  Leauthaud and Fabian Schmidt for providing the results from shear and magnification used in Schmidt et al.~2012 and shown in Fig.~\ref{fig:sigma2}.
EG is supported by spanish MINECO  grants PGC2018-102021-B-100 and EU grants LACEGAL
734374 and EWC 776247 with ERDF funds.
IEEC is funded by the CERCA program of the Generalitat de Catalunya. 
SJS and JAT acknowledge support from DOE grant DE-SC0009999 and NSF/AURA grant N56981C.
Part of this work was performed under the auspices of the U.S. Department of Energy by Lawrence Livermore National Laboratory under Contract DE-AC52-07NA27344.

\section*{Data Availability Statement}

The data and codes used in this article will be shared on reasonable request to the corresponding author.

\end{document}